\documentclass[12pt]{iopart}

\usepackage{iopams}  
\usepackage{graphicx}

\begin{document}

\title[Topological Approach to Microcanonical Thermodynamics]{Topological Approach to Microcanonical Thermodynamics and Phase Transition of Interacting Classical Spins}

\author{F A N Santos$^{1,2},$ L C B da Silva$^{1,2}$ and M D Coutinho-Filho$^{2}$}

\address{$^1$ Departamento de Matem\'atica, Universidade
Federal de Pernambuco, 50670-901, Recife-PE, Brazil}
\address{$^2$ Laborat\'orio de F\'{\i}sica Te\'orica e
Computacional, Departamento de F\'{\i}sica, Universidade Federal de
Pernambuco, 50670-901, Recife-PE, Brazil}
\ead{fansantos@dmat.ufpe.br}
\begin{abstract}
We propose a topological approach suitable to establish a connection between thermodynamics and topology in the microcanonical ensemble. Indeed, we report on results that point to the possibility of describing {\it interacting classical spin systems} in the thermodynamic limit, including the occurrence of a phase transition, using topology arguments only. Our approach relies on Morse theory, through the determination of the critical points of the potential energy, which is the proper Morse function. Our main finding is to show that, in the context of the studied classical models, the Euler characteristic $\chi(E)$ embeds the necessary features for a correct description of several magnetic thermodynamic quantities of the systems, such as the magnetization, correlation function, susceptibility, and critical temperature. Despite the classical nature of the studied models, such quantities are those that do not violate the laws of thermodynamics [with the proviso that Van der Waals loop states are mean field (MF) artifacts]. We also discuss the subtle connection between our approach using the Euler entropy, defined by the logarithm of the modulus of $\chi(E)$ per site, and that using the {\it Boltzmann} microcanonical entropy. 
Moreover, the results suggest that the loss of regularity in the Morse function is associated with the occurrence of unstable and metastable thermodynamic solutions in the MF case. The reliability of our approach is tested in two exactly soluble systems: the infinite-range and the short-range $XY$ models in the presence of a magnetic field. In particular, we confirm that the topological hypothesis holds for both the infinite-range ($T_c \neq 0$) and the short-range ($T_c = 0$) $XY$ models.
Further studies are very desirable in order to clarify the extension of the validity of our proposal.
\end{abstract}

\maketitle

\section{Introduction}

Much effort has been devoted to correlate the underlying topological and geometrical properties of equipotential manifolds in phase space with the occurrence of a phase transition (PT) at a critical value of the energy $E_c$ \cite{Pettinilivro,KastnerRMP2008}. Indeed, for a class of confining short-range potentials, very strong arguments support the proposal that a topology change of the configuration space should take place as the system undergoes a finite temperature PT \cite{Pettinilivro,FranzosiPRL2004}. The signature of the referred topology change is expected to be printed in the topological invariants of the manifold, such us the Euler characteristic $\chi(E)$; besides, the arguments do not exclude mean field (MF) models. Moreover, the relevance of the singularities associated with the stationary points (critical points) of the potential energy has been emphasized \cite{KastnerPRL2007} by the following condition: at a PT the density of Jacobian's critical points diverges in the thermodynamic limit, or, by the same token, the determinant ($D$) of the Hessian matrix of the potential should be asymptotically flat at the transition. The two above-mentioned conditions were shown to be fulfilled in several MF models: $XY$ and k-trigonometric \cite{KastnerPRL2007}; $XY$ on $AB_2$ chains under frustration- or field-induced PT's \cite{SantosPRE2009,AB2chains}; and in a model of self-gravitating particles \cite{NardiniPRE2009}. In addition, it has been proposed \cite{SantosPRE2009} that in the mentioned models the following property holds: the simultaneous occurrence of the two necessary conditions, namely the $D$-flatness condition and the discontinuity or cusp-like pattern exhibited by $\chi(E)$ at $E_c$, emerges as a necessary and sufficient condition for the occurrence of a finite temperature PT.
However, it was shown \cite{KastnerPRL2011} that, contrary to previous results \cite{FranzosiPRL2000}, the two-dimensional (2d) short-range $\phi^4$ model with O(1) symmetry, which is in the same universality class of the 2d Ising model, violates the theorem proposed in  \cite{FranzosiPRL2004}: neither the transition at $E_c$ is printed in the Euler characteristic nor the Hessian determinant is flat; similar results have been found in numerical simulations of the 2d- and 3d short-range $XY$ models \cite{NerattiniPRE2013}. Very recently, this counterexample was circumvented by an extension \cite{theo} of the first version of the theorem \cite{FranzosiPRL2004}. In fact, it was shown \cite{theo} that the phase transition of the $\phi^{4}$ model stems from an asymptotic ($N \rightarrow \infty$) change of topology of the energy level sets, in spite of the absence of critical points of the potential energy corresponding to a PT. Therefore, the main idea underlying the topological hypothesis is preserved: a PT might correspond to a change in the topology of the manifolds whose geodesics define the motions of the system \cite{PRL1997}.

We also mention that for discrete models, such as the 1d \cite{RehnBJP2012} and 2d- Ising and $q\mbox{-states}$ Potts \cite{Blanchard} models, topological arguments can also contribute to the understanding of the thermodynamic PT of the model \cite{congressoStanley70}.

In this work, we shall extend the above-mentioned ideas and propose a topological approach to establish a connection between thermodynamics and the topology of configuration space. Indeed, we shall report on results that point to the possibility of using only topology arguments in order to describe the microcanonical thermodynamics of {\it interacting classical spins} in the thermodynamic limit, including the occurrence of a PT. Our approach relies on Morse theory \cite{Milnor,Matsumoto}, through the determination of the critical points of the potential energy, which is the proper Morse function. Our main finding is to show that, in the context of the studied classical spin models, the Euler characteristic embeds the necessary features for a correct description of magnetic quantities, such as the magnetization, susceptibility, and critical temperature. Despite the classical nature of the studied models, such quantities are those that do not violate the laws of thermodynamics (with the proviso that Van der Waals loop states are mean field artifacts). We also discuss the subtle connection between our approach using the Euler entropy, defined by the logarithm of the modulus of $\chi(E)$ per site, and that using the Boltzmann microcanonical entropy. The reliability of our approach is tested in two systems exactly soluble by standard methods of statistical mechanics: the MF-XY and the 1d short-range XY models.

\section{Euler characteristic and the microcanonical ensemble} 

The problem of geometrical and topological properties of the microcanonical entropy associated with a standard classical Hamiltonian with many degrees of freedom $H=\sum_i^n p_i^2/2m+V(q)$, where $q = (q_1,...,q_n)$ and $p = (p_1,...,p_n)$, has been undertaken in detail in  [1]. In particular, it has been established that a singular behavior of thermodynamic observables is originated only from a contribution of topological origin. These studies gave rise to the so-called {\it topological hypothesis}, which suggests a change of the measure and topology of the equipotential manifold during a phase transition. In fact, it was suggested that the Boltzmann entropy per independent degree of freedom (large $N=2n-1$), $S(E)=\frac{k_B}{N}\ln\Omega_N(E)$, could possibly be approximated by the addition of a topological contribution and a smooth function of $E$, in the following alternative forms
\cite{CasettiJSP2003,AngelaniPRE2005,Pettinilivro}: 
\begin{equation}
 S(E)\approx \frac{1}{N}\ln N_c(M_E)+\mathcal{R}(E)\,\,,\label{eq::approxEntByNc}
\end{equation}
and
\begin{equation}
S(E)\approx \frac{1}{N}\ln\vert \chi(M_{E})\vert+\mathcal{R}(E)\,\,,\label{eq::approxEntByChi}
\end{equation}
where $M_E=\{q\in M\,:\,V(q)/N\leq E\}$, $N_c(M_E)=\sum_{k=0}^N\mu_k(M_E)$ is the total number of critical points lying in $M_E$, $\chi(M_E)=\sum_{k=0}^N(-1)^k\mu_k(M_E)$ is the Euler characteristic, $\mu_{k}(M_{E})$ is the Morse number, which counts how many critical points of index $k$ lie in $M_{E}$, and $\mathcal{R}(E)$ is analytical (or at least $C^{2}$) around the transition point. Notice that, while (\ref{eq::approxEntByNc}) does not seem to present any mathematical inconsistency as an entropy, the fact that the Euler characteristic of the associated equipotential manifold could be zero in some cases deserves special care, and will be discussed in Section II.B.

We remark that, since $\Omega_N(E)=\frac{1}{N!}\int_{\Sigma_E} \Vert\nabla H\Vert^{-1}\rmd\sigma$, where $\Sigma_E$ is the constant-energy hypersurface in the $2n$-dimensional phase space, the derivation \cite{Pettinilivro} of the approximants for the entropy in  (\ref{eq::approxEntByNc}) and (\ref{eq::approxEntByChi}) were made using the Gauss-Bonnet-Hopf Theorem and a complementary result due to Chern and Lashof: $\int_{\Sigma_E}\,\rmd\sigma\,K_G=\frac{1}{2}\mbox{vol}(\mathbb{S}_1^{N-1})\chi(\Sigma_E)$ and $\int_{\Sigma_E}\,\rmd\sigma\,\vert K_G\vert\geq \frac{1}{2}\mbox{vol}(\mathbb{S}_1^{N-1})\sum_{k=0}^{N-1} b_k(\Sigma_E)$, respectively, where $\mathbb{S}_1^{N-1}$ is an $(N-1)$-dimensional sphere of unit radius, $K_G$ is the Gauss-Kronecker curvature of $\Sigma_E$, and $b_k(\Sigma_E)$ is the $k$-th Betti number of $\Sigma_E$. In addition, it is argued that, for large $N$, $\Sigma_E$ concentrates on $\mathbb{S}_{\langle 2K\rangle^{\frac{1}{2}}}^{n-1}\times M_{v=\langle V\rangle}$, where $\mathbb{S}_{\langle 2K\rangle^{\frac{1}{2}}}^{n-1}=\{p:\sum_{i=1}^np_i^2=\langle 2K\rangle\}$ and $M_{v=\langle V\rangle}=\{q:V(q)\leq\langle V\rangle\}$, and also that $\mu_i(M)\approx b_k(M)$. In fact, an alternative derivation of (\ref{eq::approxEntByNc}) \cite{Pettinilivro} is to use the condition valid for ``normal'' systems in the sense of statistical thermodynamics \cite{Kubo}, namely that the Boltzmann entropy is well approximated by the logarithm of the total number of microscopic states $\Omega_0(E)$:
\begin{equation}
 \Omega_0(E)=\frac{1}{N!}\int_{H(p,q)\leq E}\,\rmd p\,\rmd q\,\,.\label{eq::defOmega0}
\end{equation}
We also emphasize that (\ref{eq::approxEntByNc}) was derived using mathematical arguments that avoid critical points, i.e., only points in energy intervals between critical values of $V(q)$ are considered. Notwithstanding, it is argued that the disjoint union of all these open sets should define, for large $N$, a good approximation for a smooth function, i.e., the entropy \cite{Pettinilivro}.  

Many authors took advantage of the expectation that the topology contribution should dominate the entropy around a finite-temperature phase transition \cite{Pettinilivro,KastnerRMP2008,SantosPRE2009,RehnBJP2012,congressoStanley70,CasettiJSP2003,AngelaniPRE2005}. In fact, the computation of the first term in  (\ref{eq::approxEntByNc}) and (\ref{eq::approxEntByChi}) became an efficient means to validate the topological hypothesis, in which case the referred quantities exhibit a cuspid, or discontinuity, as the energy crosses the critical value $E_c$. In particular, for the MF-$XY$ model in a field the ferromagnet transition is absent, thereby leading to a monotonically decreasing contribution of the Euler characteristic to the entropy for $E>E_c$, i.e., 
$\frac{\rmd}{\rmd E}(\lim_{N\to\infty}\frac{1}{N}\ln\vert \chi(M_E)\vert)<0$,
while the contribution of the total number of critical points saturates, i.e., $\rmd\tau(E)/\rmd E=0$, where [notice that in  [1] $\tau(E)=\tilde{S}(E)$]
\begin{equation}
\tau(E)=\frac{1}{N}\ln N_c(M_E)\,.\label{def::approxEntByNc}
\end{equation}
In short, in the thermodynamic limit, the two referred contributions are identical for $E<E_c$, monotonically increasing functions of the energy, and bona-fide contribution to the entropy, whereas for $E> E_c$ they differ drastically and claims for a careful interpretation.


\subsection{Euler Topological Approach to Microcanonical Thermodynamics: Interacting Classical Spin Models}

Here, we shall built on these ideas and results to propose a topological approach that appears suitable to describe the thermodynamics and phase transition of {\it interacting classical spins} . These models are a restricted set of classical systems, since their Hamiltonian is defined by the spin interaction potential only, without any sort of kinetic term. Notwithstanding, as shown below, it is gratifying that some soluble models that have been used to test the topological hypothesis, such as the MF-XY model \cite{CasettiJSP2003}, 1d $XY$ short-range model \cite{CasettiJSP2003}, and the k-trigonometric model \cite{AngelaniPRE2005}, do belong to this class of models and can be successfully analyzed within our framework. Our topological approach is based on the integration with respect to Euler characteristic, the $\chi$-integral, as proposed by Viro \cite{Viro} and Schapira \cite{cha}, and corresponds to an analytical interpretation of the classical Euler characteristic. Despite that in the referred proposal \cite{Viro,cha}, the Euler characteristic as a measure has been devised in a context distinct from that of differentiable manifolds, a Morse-theoretic interpretation in the context of manifolds was given to the corresponding $\chi$-integral \cite{Baryshnikov}, and also applied to object enumerations in networks \cite{bell}. In this work, these ideas and methods are used in the context of equipotential manifolds associated with systems of interacting classical spins in the microcanonical ensemble.

{\it Euler integration} \cite{Viro,cha,Baryshnikov}: Let $V(q)$ be a Morse function in a $N$-dimensional manifold $M$. Denote by ${\mathcal C}(V)$ the set of critical points $q_{c}$ of $V(q)$. For each $q_{c}$ the index, $k(q_{c})$, is defined as the number of negative eigenvalues of the Hessian of $V$ at $q_{c}$. Then, the integral of a function $f(q)$ over $M$ with respect to the Euler (Poincar\'e) characteristic, i.e., the $\chi$-integral, is defined by
   \begin{equation}
       \int_{M} f\, \rmd\chi=\sum_{q_{c}\,\in\,\mathcal{
C}(V)}(-1)^{k(q_{c})}f(q_{c}),\label{def::ChiIntegral}
   \end{equation}
which implies
   \begin{equation}
       \int_{M_{E,\delta E}} \rmd \chi =
\chi(M_{E,\delta E})=\sum_{k=0}^{N}(-1)^{k}\mu_{k}(M_{E,\delta E}),\label{def::EulerChar_Fatia}
   \end{equation}
where the Morse number $\mu_{k}(M_{E,\delta E})$ counts how many critical points of index $k$ lie in $M_{E,\delta E}=\{q \in M\,:\,\vert V(q)/N -E\vert\leq\delta E\}$; and also
\begin{equation}
\chi(M_{E})=\int_{M_E}\,\rmd\chi=\sum_{k=0}^{N}(-1)^{k}\mu_{k}(M_{E})\label{def::EulerChar}.
   \end{equation}
The motivation to introduce $\delta E$ will become clear below. The $\chi$-integral is the main mathematical concept underlying our topological approach; we also emphasize that this integral is neither a Riemann nor a Lebesgue integral.

It will prove very useful to define the Euler entropy
\begin{equation}
      S_{\chi}(E)=\frac{1}{N}\ln\vert\chi(M_{E,\delta E})\vert,
      \label{equivBetweenEntropyANDSchi}
\end{equation}
%
in analogy [see also  (\ref{eq::approxEntByNc}) and (\ref{eq::approxEntByChi})] with the microcanonical Boltzmann entropy related to $V(q)$ ($k_{\mathrm{B}}\equiv1$):
\begin{equation}
 S(E)\equiv\frac{1}{N}\ln W(E,\delta E)\,\,,\label{def::BoltzEntropy}
\end{equation}
where
\begin{equation} 
 W(E,\delta E)=\displaystyle\int_{M_{E,\delta E}}\,\rmd q\cong\Omega(E)\,\delta E\,\,;\label{eq::PesoTermo}
\end{equation}
in the above equation, $W(E,\delta E)$ is the thermodynamic weight, $\Omega(E)$ is the density of states, and $\delta E$ is a small allowance in energy of no relevance in the thermodynamic limit ($\delta E\to0$ for classical systems) \cite{Gallavotti}.
 
Further, the analogy with the mean value of a thermodynamic quantity $\mathcal{O}$, defined by
   \begin{equation}
\langle\mathcal{O}\rangle(E)=W(E,\delta E)^{-1}\int_{E<E(q)<E+\delta E}\mathcal{O}(q)\,\rmd q,\label{def::meanValue_thermodyn}
   \end{equation}
suggests that, in the $\chi$-integral context, $\langle\mathcal{O}\rangle$ reads:
   \begin{equation} \label{vm}
\langle\mathcal{O}\rangle_{\chi}(E)=\chi(M_{E,\delta E})^{-1}\displaystyle\sum_{\{q_c\,:\,\vert E-E(q_{c})\vert\leq\delta E\}}(-1)^{k(q_{c})}\mathcal{O}(q_{c}),
   \end{equation}
where in (\ref{vm}) we emphasize that the degeneracy of the critical points must be properly considered. For example, in MF models the degeneracy is equal to the Morse number, whereas for the short-range XY model the counting is a bit more complex.
Last, we introduce the Euler temperature $(T_{\chi})$ through
the analogy
   \begin{equation}\label{eulert}
      \frac{1}{T}=\frac{\partial S}{\partial E} \rightarrow
\frac{1}{T_{\chi}}=
\lim_{N\rightarrow\infty}\frac{S_{\chi}(E_2)-S_{\chi}(E_1)}{E_2-E_1}
   \end{equation}
In computing the thermodynamic temperature, we can make $\Delta E=E_2-E_1$ arbitrarily small. However, in computing $T_{\chi}$, we should pay attention to the Noncritical Neck Theorem \cite{Palais}, and therefore, we choose $\Delta E$ as the exact distance between neighboring critical values $E_2$ and $E_1$, otherwise the Euler characteristic would not change. This choice is adequate since we expect that the distance between neighboring critical points approaches zero in the thermodynamic limit. This derivation process also applies to any $E\mbox{-function}$ topological invariant. We stress that $T_{\chi}$ is an intensive variable in the context of the topological approach. However, due to the classical nature of the models, the Euler temperature is identical to the thermodynamic one only in some special limits: at $T=0$ and at $T=T_c$. 

Finally, in order to apply the techniques introduced in this Section, some conditions should be satisfied. We believe that a minimal list of necessary requirements would be: (i) the critical levels of the potential $V(q)$ are distributed along the same energy interval where the Boltzmann entropy is defined; and (ii) the distance between neighboring critical levels approaches zero as $N\to\infty$, i.e., the set of these levels should be densely distributed in the thermodynamic limit. 

As previously mentioned in the Introduction, the 2d short-range $\phi^4$ model with O(1) symmetry was shown \cite{KastnerPRL2011} to  violate the necessity theorem proposed in  \cite{FranzosiPRL2004}: the transition at $E_c$ is not printed in the Euler characteristic.  From this, it is clear that our proposal does not apply to the 2d $\phi^4$ model, since condition (i) above is not satisfied. Interestingly, we have noticed that other models, such as the MF $\phi^4$ model \cite{AndronicoPRE2004}, the 1d Peyrard-Bishop model \cite{GrinzaPRL2004}, and the Burkhardt solid-on-solid model \cite{AngelaniPRE20051dModels}, where the behavior of the Euler characteristic is in conflict with the topological hypothesis, have a common feature, namely the configuration space is non-compact. In fact, for all the models mentioned above, the configuration space is $\mathbb{R}^N$, which is clearly non-compact. Whether the condition of compactness is the missing hypothesis in the necessity theorem \cite{FranzosiPRL2004} is an open question. Lastly, at the MF level, we remark that for all models satisfying the topological hypothesis, a consistent Landau $\phi^4$ model near $E_c$ can be derived by requiring that the associated order parameter is small (scaling region), with critical exponents obeying scaling relations.


\subsection{Boltzmann and Euler Entropies, and Negative Spin Temperature}

As previously stated, it is well known that for normal systems in the sense of statistical thermodynamics \cite{Kubo} $W(E,\delta E)$ can be well approximated by $\Omega_0(E)$. However, in ideal spin systems with an energy upper bound (no positive unbounded kinetic energy term), high energy states may violate this condition and give rise to negative spin temperatures. Recently, this feature raised some controversy \cite{BraunScience2013,DunkelNaturePhys2014} and in the following we shall digress on this matter.

In an attempt to resolve the discrepancy between the results derived using the Boltzmann entropy and $\Omega_0(E)$ [for the referred classical models], in  \cite{DunkelNaturePhys2014} the authors suggested that the correct approach is the one using $\Omega_0(E)$, thereby excluding the occurrence of states with negative spin temperature and replacing them by a state of infinite temperature in the entire high energy region associated with the saturation of $\Omega_0(E)$. However, in this work we shall present statistical mechanics and topology arguments that point to the consistency of the Boltzmann and Euler entropies in the description of the {\it magnetic} properties of the infinite and short-range $XY$ models, in which case a proper interpretation of negative spin temperature states is provided in the context of ideal interacting classical spin systems.

Some remarks are in order: 

(i) Since, to the best of our knowledge, models suitable to a topological approach based on Morse theory behave classically, some well known drawbacks due to the continuous energy spectrum may arise, particularly at low temperatures, thereby giving rise to results that might violate the laws of thermodynamics, such as a nonzero specific heat. Most importantly, in the case of the studied $XY$ models, the Boltzmann entropy exhibits singular behavior as $T\to0$ \cite{RuffoPRE1995,Mehta}. On the other hand, the Euler entropy is positive definite and, as such, it approaches zero as $T\to0$ (and the specific heat as well). Remarkably, despite the referred subtleties due to the classical nature of the models, their magnetic properties, such as the mean magnetization, susceptibility, critical temperature, and correlation function (short-range model), do not violate the laws of thermodynamics (with the proviso that Van der Waals loop states are mean field artifacts) and can be exactly calculated using {\it either} the Boltzmann description of microcanonical thermodynamics or the topological approach based on the Euler entropy and Morse theory. 

(ii) From the above discussion, it is clear that the classical nature of the models precludes the possibility of an equality between the Boltzmann and Euler entropies  in the thermodynamic limit, as for the 1d Ising model in a field \cite{RehnBJP2012}, in which case the equality follows from the discrete symmetry of the model. In fact, using a topological procedure to study phase transitions that suits discrete models \cite{Blanchard}, it was shown that, in zero field,
\begin{equation}
\lim_{N\to\infty}\frac{\ln\vert\chi(_{XY})\vert}{N}=\lim_{N\to\infty}\frac{\ln\vert\chi(_{Ising})\vert}{N}=\lim_{N\to\infty}\frac{S_{Ising}}{N},
\end{equation}
where the first and second terms are the per-site Euler entropies of the 1d- short range $XY$ and Ising models, respectively, while the last one is the Boltzmann entropy of the 1d Ising model. Moreover, there is no violation of the laws of thermodynamics at low temperatures, although the high-energy states do exhibit negative spin temperatures, which are formally mapped onto antiferromagnetic states at positive temperatures. This feature is also confirmed in this work, in the context of the MF and the 1d short-range $XY$ models.

(iii) From a statistical mechanics viewpoint, the equivalence between the canonical and microcanonical ensembles implies the following relations in the thermodynamic limit ($N\to\infty$): $Z(\beta)=\int_{E_{0}}^{\infty}\rme^{-N\beta E}\Omega (E)\, \rmd E$ and $\Omega(E)=\frac{1}{2\pi \rmi}\int_{\beta'-\rmi\infty}^{\beta'+\rmi\infty}Z(\beta)\rme^{N\beta E}\,\rmd\beta,$ with $\beta$ defined in the complex plane and $E_0$ is the minimum energy of the system. It is well known that, for $N \to\infty,$ the use of saddle point techniques, in either of the two previous relations, enables us to derive the thermodynamic relation connecting both ensembles: $F=E-TS,$ where $F=-\frac{1}{\beta N}\ln Z$ and $S=\frac{1}{N}\ln\Omega$ are the Helmholtz free energy and the Boltzmann entropy per degree of freedom \cite{khinchinANDkubo,Kubo}, respectively. Notwithstanding, here it is shown that in order to achieve a more complete understanding of the equivalence of ensembles in the context of the models, it is instructive to extend the domain of $T$ to negative spin temperatures and to search for the unstable and metastable solutions. We stress that negative spin temperature states, which are ``formally'' allowed only for systems with an energy upper bound and spins loosely coupled to the other degrees of freedom (ideally uncoupled in our spin-only systems), were first considered in the context of experiments in nuclear spin systems \cite{purcel}, and subjected to an analysis of relevant aspects of their thermodynamics and statistical mechanics properties \cite{ramsey,klein}. A variety of models predicting states with negative temperatures has attracted continuous interest: anti-shielding effect and superluminal light propagation under reversed electric fields \cite{ShenScripta2003}, cosmological model describing dark energy and supermassive black holes \cite{G-DPRD2004}, unconfined quark-gluon plasma \cite{PariharPRC2006}, decoherence \cite{SchmidtPRE2005}, PTs, metastability, entanglement in bipartite quantum systems \cite{ParisiPRA2010}, and optical lattices under parabolical potentials \cite{MoskPRL2005,BraunScience2013}. However, it has been shown \cite{RomeroRochinPRE2013} that states at negative {\em absolute} temperature ($T_N$) are metastable and heat can flow irreversibly to a reservoir at an absolute positive temperature ($T_P$). In the experiments of Purcell and Pound \cite{purcel} the lattice plays the role of the $T_P$ reservoir, albeit the observed relaxation time is indeed much longer than those relaxation times for ordinary spin excited states. The referred metastability, or the lack of equilibrium $T_N$ states, is a result of the combination of Ramsey's postulate of the second law of thermodynamics for $T_N$ states, with that of Kelvin for $T_P$ states; although both postulates agree with that of Clausius: heat flows from a hot reservoir to a cold one. In fact, in the microcanonical ensemble, $T_N$ states appear as energy increases from the low-energy range to the high-energy one, i.e., from cold to hot; thereby the equivalence with the canonical ensemble would imply the following $T$-range: $0^{+},\ldots,+T,\ldots,\infty,-\infty,\ldots,-T,\ldots,0^{-}$. Despite that our main goal in this work is to establish the connection between statistical mechanics and topology in the context of the microcanonical ensemble of macroscopic systems in equilibrium, we remark that the extensive Boltzmann entropy $S=\ln W$ is the starting point in the study of the irreversible time evolution of macroscopic systems out of equilibrium \cite{LebowitzPhysA}, as well as in the microcanonical description of the thermodynamics and phase transitions in ``small'' systems \cite{gross}. In this context, the case $N=2$ of the infinite-range $XY$ model discussed in Section IV is proved quite instructive. 

In short, in the context of ideal spin-only systems, such as the ones studied in this work, the concept of negative spin temperature is a useful one in the analysis of high energy states, which, as for the 1d Ising model in a field \cite{RehnBJP2012}, are formally identified as antiferromangnetic states at positive spin temperature. Therefore, with this proviso, the nature of the spin configuration states will be analyzed under the thermodynamic stability conditions related to the magnetization of the system only.

(iv) Now we discuss the possibility of a vanishing value for the Euler characteristic, that would lead to a singular Euler entropy. It follows from the well known Poincar\'e Duality Theorem \cite{Matsumoto} that under the assumptions of connectedness, orientability and closedness (i.e., compact and without boundary \cite{Matsumoto}), the Euler characteristic of an odd-dimensional manifold is zero. However, this result cannot be applied to the sets used in our work, in the context of  (\ref{def::EulerChar_Fatia}) and (\ref{def::EulerChar}), since $M_E$ and $M_{E,\delta E}$ have boundaries and therefore the hypotheses of the theorem are not met. Indeed, the hypothesis of closedness can not be omitted in our topological approach suitable to describe interacting classical spin systems exhibiting a PT. For example, if $M_E$ is obtained by the attachment of the minima points, i.e. when we pass the first critical value of $V(q)$, say $E_{\min}$, then $\chi(M_{E_{\min}})=\displaystyle\sum_{k=0}^N(-1)^k\mu_k(M_{E_{\min}})=\#\mbox{ minima}>0,$ since $\mu_k=0$ for $k>0$ and $\mu_0=\#\mbox{ minima}$, regardless of the parity of the dimension $N$. In addition, if one invokes an analogy with the density of states $\Omega(E)$, in (\ref{eq::PesoTermo}), then we would define $S_{\chi}$ through the Euler characteristic of $\partial M_E=\{q \in M\,:\,V(q)/N=E\}$, where $E$ is a critical level of $V(q)$. These level sets are compact, have no boundary, and are of dimension $N-1$. Therefore, the application of the Poincar\'e Duality Theorem would imply $\chi(\partial M_E)=0$ for $N$ even. However, when $E$ is a critical level of $V(q)$, the set $\partial M_{E}$ is not a manifold, since the neighborhood of a critical point $p\in\partial M_E$ is a degenerate quadric, which do not qualify as a manifold (see page 255 of  [1]). In this case, we believe that a proper way to calculate $\chi(\partial M_E)$ is to use $M_{E,\delta E}$ and to take $\lim_{\delta E\to0}\chi(M_{E,\delta E})$, in which case Morse theory can be applied, as described in Sections 2.1, 2.4, and 2.5 (in particular, see the case $N=2$ discussed in Section 4.1). Besides, we also remark that in the spin-only models analyzed so far \cite{SantosPRE2009,CasettiJSP2003,AngelaniPRE2005}, the Euler characteristic does not vanish for critical energies in the interval $E_{\min}<E<E_{\max}$, regardless the parity of $N$. 

In the next sections, we shall focus on the use of the Euler and the microcanonical Boltzmann entropies, through  (\ref{equivBetweenEntropyANDSchi}) and (\ref{def::BoltzEntropy}), in order to describe the thermodynamics and PT of the MF and the 1d short-range $XY$ models. The results will allow us to comment on the connection between thermodynamics and topology, including subtle aspects that appear from the specific methodologies of the approaches and the classical nature of the models.

\section{Equivalence of Ensembles in the Infinite-range $XY$ model} 
We shall first use the above proposed topological approach to describe the infinite-range $XY$ model in the presence of a field defined by the following potential energy of $N$ ferromagnetically coupled classical spin vectors of fixed length ($\equiv 1$) \cite{RuffoPRE1995}:
   \begin{equation} \label{Pot_energy}
V(\theta;h)=\frac{1}{2N}\sum_{i,j=1}^{N}[1-\cos(\theta_{i}-\theta_{j})]-h\sum_{i=1}^{N}\cos\theta_{i},
   \end{equation}
where $\theta =(\theta_{1},\ldots,\theta_{N})$, with $\theta_i \in[0,2\pi)$ being the position (angle) of the $i$-th spin with $h$ in the $x$ direction, i.e., the $x$-axis of the chosen $xy$ reference axes. In fact, due to the rotational invariance symmetry and infinite-range interactions of the model, the chosen axes and the spin positions are immaterial \cite{SantosPRE2009}.
It is convenient to introduce the magnetization, given by
   \begin{equation}
   \label{mag} \mathbf{m}(\theta)\equiv
   (m_{x}(\theta),m_{y}(\theta))=\Big(\frac{1}{N}\sum_{i=1}^{N}\cos\theta_{i},
   \frac{1}{N}\sum_{i=1}^{N}\sin\theta_{i}\Big).
   \end{equation}
Therefore, the potential energy per spin for a given configuration
$\theta$ reads:
  \begin{equation}\label{energy_per_spin}
E(\theta;h)=\frac{V(\theta;h)}{N}=\frac{1}{2}(1-m_x^2-m_y^{2})-hm_{x}.
 \end{equation}
 
\par The critical points of $E(\theta;h)$ are obtained by solving the equation
\begin{equation}\label{first_derivate}
 \frac{\partial E(\theta;h)}{\partial \theta_{i}}= m_{x}\frac{\sin
\theta_{i}}{N}-m_{y} \frac{\cos
\theta_{i}}{N}+h\frac{\sin\theta_{i}}{N}=0,
\end{equation}
for which it is only pertinent to consider the solutions $\theta_i=0$ or $\pi$, and $m_y=0$ (rotational invariance), which correspond, in the thermodynamic limit, to all values of $m_x \equiv M \in [-1,1]$; besides, the critical point $M=-h$ is governed by an external continuous parameter (the field $h$ conjugate to $M$). However, for a fixed value of $h$, some values of $M$ shall manifest more significantly, both in the thermodynamic description and in the $\chi$-topological approach. We shall thus consider the $E(\theta;h)$ {\em vs.} $M$ curves, illustrated in figure 1(a) for $h = 0, \pm 0.25, \pm 0.5, \pm 0.75 \pm 1.0$, and $\pm 1.5$, in light of the special critical points discussed below. In zero field, $\partial^{2} E(\theta;0)/\partial \theta_{i}^{2}=
M\cos \theta_{i}/N-1/N^2$. Therefore, at the trivial critical point $M=0$,
$E(\theta_c;h=0)$ attains the maximum value: $E_{\max}(h=0)=\frac{1}{2}$ (magenta square), in which case both solutions $\theta_i=0$ and $\theta_i=\pi$ occur with equal probability at each site $i=1,\ldots,N$. Further, at the critical point $\theta_i=0$
 ($\theta_i=\pi$) $\forall i$, i.e., $M=1$ ($M=-1$), $E(\theta_c;h=0)$ attains the minimum value: $E_{\min}(h=0)=0$ (red triangles).
However, in the presence of a field, $\partial^{2}
E(\theta;h)/\partial \theta_{i}^{2}=(M+h)\cos \theta_{i}/N-1/N^2$;
hence, for $h>0$ ($h<0$) $E(\theta_c;h)$ attains the minimum value,
$E_{\min}(h)=-|h|$ (red triangles), only at the critical point $M=1$ ($M=-1$).
 Further, for $0<|h|\leq 1$
  the maximum of $E(\theta_c;h)$  is shifted to the critical point
  \begin{equation}
   M=-h,  \mbox{ } E_{\max}(h)=\frac{1}{2}+\frac{h^2}{2},
  \end{equation}
illustrated by the magenta dashed line and squares at distinct field values.
Last, for $h \geq 1$ ($h \leq -1$), $E(\theta_c;h)$ attains the maximum value $E_{\max}(h)=|h|$ at the critical point $M=-1$ ($M=1$), illustrated by red squares. We also remark that for $0< h < 1$ ($-1 <h < 0$), $M=-1$ ($M=1$) is a critical point at which $E(\theta_c;h)$ has the value $E(h)=|h|$ (curves without symbols at $M=\pm 1$), which is neither a maximum nor a minimum value [saddle point solutions in figures 3(b)-(d)]. The above analysis of the critical point solutions of $E(\theta_c;h)$, namely $M=0,\pm 1$, and $-h$, will prove most relevant in the understanding of the thermodynamics and phase transition exhibited by the model, particularly in the topological context \cite{SantosFigE}.

\begin{figure*}[t]
    \centering{
     \includegraphics[width=\linewidth]{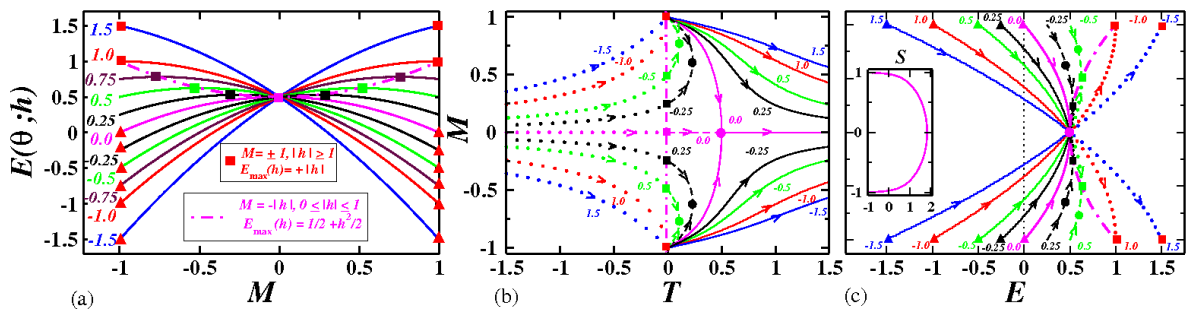}}
    \caption{Infinite-range $XY$ model: (a) potential energy per spin $E(\theta;h)$ {\em vs.} $M$. (b) $M$ {vs.} $T$ canonical diagram. (c) $M$ {vs.} $E$ microcanonical diagram. The full (dotted) lines correspond to stable solutions with $T>0$ ($T<0$); the dashed lines are the $M$-$T$ van der Waals loops for $h=\pm 0.25$ and $\pm 0.5$, with spinodal points at $T_{max}(h)$ (full balls).}
\end{figure*} 
\begin{figure*}[tbp]
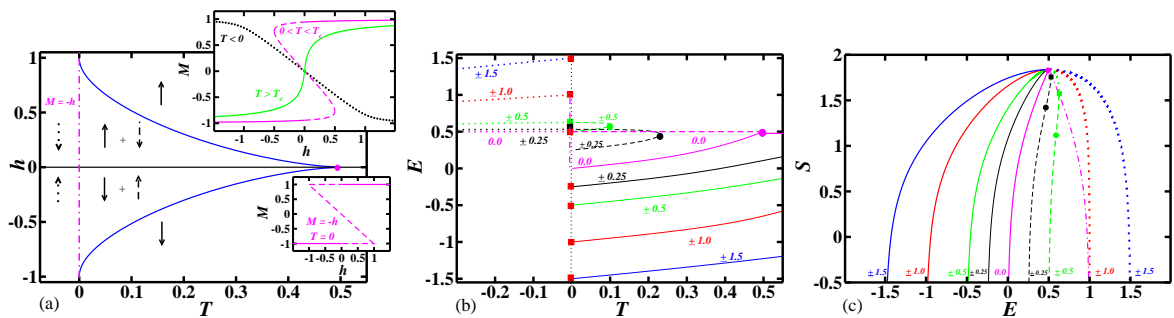

\centering
    {\includegraphics[width=0.33\linewidth]{Fig2a.eps}}
    {\includegraphics[width=0.32\linewidth]{Fig2b.eps}}
    {\includegraphics[width=0.32\linewidth]{Fig2c.eps}}
          \caption{(Colour online). (a) $h$ {\em vs.} $T$ phase diagram. In the upper (lower) inset we illustrated the $M$ {\em vs.} $h$ diagram for fixed values of $T$ ($T=0$). (b) $E$ {\em vs.} $T$ diagram. (c) $S$ {\em vs.} $E$. The $T>0$ ($T<0$) stable solutions  with $\frac{\partial^2 S }{\partial E^2}<0$ are illustrated by full (dotted) lines; the metastable (unstable) solutions satisfy  $\frac{\partial^2 S }{\partial E^2}<0$ ($\frac{\partial^2 S}{\partial E^2}>0$) and are  illustrated by dashed lines. At the spinodal points [blue curves in figure 2(a) and full balls in figures. 2(b) and 2(c)], $\frac{\partial^2 S }{\partial E^2}|_{T_{\max}}=0$ . The dash-dot line corresponds to $S$ {\em vs.} $E_{max}(h)$ for $0\leq h \leq 1$.}
\end{figure*}

\subsection{Canonical Ensemble}
\par The exact solution of the model in the canonical ensemble is obtained by the computation of the partition function, $Z(\beta \equiv
1/T,h,N)=\int\limits_{0}^{2\pi}\prod\limits_{i=1}^{N}\rmd\theta_{i}\exp[-N\beta
E(\theta;h)]$. Performing the integration over the $N$ angular variables and using the saddle point method ($N\rightarrow \infty$), the free energy per particle, $F( z;\beta,h)=-\lim\limits_{N \rightarrow
\infty}\frac{1}{\beta N}\ln Z(\beta,h,N)$, reads \cite{SantosPRE2009}:
      $   F( z;\beta, h)=\frac{1}{2}+\beta^{-1}\Big[|
z|^{2}/2\beta-\ln2\pi I_{0}(z+\beta h)\Big],$
  where $I_{n}$ is the $n$-order modified Bessel function and $z$ is the solution of the saddle point self-consistency equation \cite{SantosPRE2009}:
   \begin{equation}\label{self_consist}
      \frac{z}{\beta}=\frac{I_{1}}{I_{0}}(z+\beta h)=M(z;\beta,h),
   \end{equation}
with magnetization $M(z;\beta,h)=-\frac{1}{\beta}\frac{\partial
F(z;\beta,h)}{\partial h}$. In zero field, the solution of (\ref{self_consist}) is $z=0$ for $\beta<\beta_{c}=2$, corresponding to a vanishing magnetization, $z\neq 0$ for $\beta>\beta_{c}$, corresponding to an ordered phase, and $\lim_{\beta\to\infty}z=\infty.$ The energy per particle, $E(z;\beta,h)=-\frac{\partial}{\partial \beta}[\beta F(z,\beta,h)]$
is given by
\begin{equation}\label{micro_energy}
 E(z;\beta,h)=\frac{1}{2}[1-M(z;\beta,h)^{2}]-hM(z;\beta,h),
\end{equation}
whose formal similarity with (\ref{energy_per_spin}) is due to mean field (MF) character of the solution, and will manifest explicitly in the microcanonical ensemble [see figures 1(a) and 1(c)]. Indeed, by inverting (\ref{micro_energy}) we find that 
   \begin{equation}
   M(E;h)=-h\pm\sqrt{h^2-2(E-\frac{1}{2})}\,\,,\label{eq::Mthermo_pm}
   \end{equation}
   which will be useful in our topological analysis. Moreover, using ~(\ref{self_consist}) and (\ref{micro_energy}) we can compute the diagrams $M$ {\em vs.} $T=1/\beta$ and $M$ {\em vs.} $E$, as shown in figures $1$(b) and (c), respectively, for $h=0, \pm 0.25, \pm 0.5,\pm 1.0$, and $\pm 1.5$.


Let us now discuss the diagram $M$ {\em vs.} $T$ shown in figure 1(b). For $0\leq h< 1$, we find three sets of solutions to ~(\ref{self_consist}) \cite{footnoteSantos}. The first,
with $M$ parallel (antiparallel) to $h$ for positive (negative) $T$, corresponds to stable solutions, and are illustrated by full (dotted) lines in figure 1(b). These stable solutions are characterized by  $\frac{\partial
M}{\partial T}<0$ ($\frac{\partial M}{\partial T}>0$) for positive
(negative) $h$.
 The second set, with $M$ anti-parallel to $h$ and $T\geq 0$ are  metastable
 solutions, such that $\frac{\partial M}{\partial T}>0$ ($\frac{\partial M}{\partial T}<0$) for positive (negative) $h$,
  as shown  for $h=\pm
 0.25$ and $\pm 0.5$ (dashed lines).
These metastable solutions start at $T=0$ with $M=\pm 1$ (red
filled squares) and only exist up to a maximum temperature,
$T_{\max}(h)$, defined by the singularity $\frac{\partial
M}{\partial T}|_{T=T_{\max}(h)}=\infty$ (full balls). The third set
of solutions, also with $M$ antiparallel to $h$ and $T\geq 0$, are
unstable solutions, such that $\frac{\partial M}{\partial T}<0$
($\frac{\partial M}{\partial T}>0$) for positive (negative) $h$.
These unstable solutions, also illustrated by dashed lines, start at
$T=0$ with $M=-h$ (filled squares)
 and meet the metastable solutions at $T_{\max}(h)$. In fact, the metastable and unstable
solutions are the corresponding van der Waals loops in the $M$ {\em
vs.} $T$ diagram \cite{footnoteSantos}.
 Notice that $T_{\max}(h=1)=0$ and for $|h|>1$ only stable solutions, emerging from $M=\pm 1$, exist in any $T$-regime.
We also remark on some special lines: i) the two $h=0$ lines meet at
 the MF critical temperature $T_{c}=\frac{1}{2}$;
 ii) the MF line of spontaneous magnetization and the $T=0$ line, have common extrema at $M=\pm 1$. Most importantly, the critical points, $M=0$, $M=\pm 1$, and $M=-h$, $-1<h<1$, form a continuum (line of critical points) along the $T=0$ axis in the interval $-1\leq M\leq 1$.
 
\subsection{Microcanonical Ensemble}   

\par We now turn to the analysis of the microcanonical phase diagram $M$ {\em vs.} $E$ shown in figure 1(c). Since $M$ and $E$ in ~(\ref{micro_energy}) depend implicitly on $T$ trough ~(\ref{self_consist}) for a given $h$, we can make a full correspondence between the canonical phase diagram in figure 1(b) and the microcanonical one in figure 1(c). Therefore, the same notation is preserved; however, one should notice that, while the arrows in figure 1(b) are oriented from the lower asymptotic temperature ($T =-\infty$) to the highest one ($T = +\infty$), in figure 1(c) the flow is defined from cold to hot, i.e., from the lower energy to the higher one, which matches the following $T$-range: $0^{+},\ldots,+T,\ldots,\infty,-\infty,\ldots,-T,\ldots,0^{-}$. Notice that the critical energy $E_c = \frac{1}{2}$ is a fixed point that corresponds to $T_c =\frac{1}{2}$ in zero field and to the asymptotic limits $T = \pm \infty$ for nonzero fields. We also remark that by having extended the $T$-range to negative spin temperatures, any physical magnetization value ($-1\leq M\leq 1$) can be accessed under a positive, a negative or a zero field $h$ (obviously, for $h\neq 0$, the $M=0$ solution is only accessed asymptotically at $T=\pm \infty $). We also mention that the solutions for opposite field directions display reflection symmetry with respect to the $T$ and $E$ axes in figures 1(b) and 1(c), respectively. Last, we have noticed that the Boltzmann entropy per particle, $S=-\frac{\partial F(z;\beta,h)}{\partial (1/\beta)}$, which can be written as
   \begin{equation}\label{entropy}
      S(z,\beta,h)=-\frac {{z}^{2}}{\beta}+\ln [2\pi I_0(z+\beta h)] -\beta
h\frac{I_{1}}{I_{0}}(z+\beta h),
   \end{equation}
can be computed as a function of $M$, with the aid of ~(\ref{self_consist}), such that  $M$ {\em vs.} $S$ displays universal behavior for any $h$  [see  inset in figure 1 (c)]. Notice that, by using the self-consistency equation, ~(\ref{self_consist}), we can infer that the thermodynamic entropy exhibits a singular behavior as $T\to0$. On the other hand, $S(z,\beta,h)\to\ln(2\pi)$ as $\beta\to\beta_c=2$.

\par In figure 2(a) we illustrate the $h$ {\em vs.} $T$ phase diagram. The blue lines are spinodal points ($T_{\max}, h_{\max})$, where the values of $T_{\max}$ are obtained from the $M$ {\em vs.} $T$ van der Waals loops in figure 1(b) for a given $h_{\max}$ defined by the $M$ {\em vs.} $h$ van der Waals loops shown in the insets for $0 \leq T < T_{c}$, including the special loop at $T=0$ defined by the critical points $M=\pm 1$ and $M=-h$, as well as other loops for $T>T_c,$ $0<T<T_c,$ and $T<0$. In the $h$ {\em vs.} $T$ phase diagram we also indicate the stable states with magnetization parallel (antiparallel) to $h$ for $T>0$ ($T<0$), which coexist with metastable and unstable states in the interior of the region defined by the spinodal lines and the line of critical points ($T=0,-1<h<1$). We mention that, for an Ising ferromagnet, a MF metastable state has a relaxation time that grows exponentially with the size of the system \cite{griff}, whereas for short-range forces the lifetime of a metastable droplet, i.e., a small region of magnetization antiparallel to the field created by fluctuations, depends critically on the radius of the droplet and the system dimensionality \cite{Chaikin,ParisiMeta}. It is also relevant to remark that van der Waals loops are of interest in studies of spinodal decomposition, interface tension and profile \cite{Fisher,gross}.

Going forward, we now examine the $T$-dependence of the energy per spin $E$, as shown in figure 2(b), whose results do not depend on the direction of $h$, including the metastable and unstable van der Waals solutions. An important feature in the $E$ {\em vs.} $T$ plot is that the $T<0$ stable solutions, for a given $|h|$, have energies higher than those corresponding to $T>0$ stable solutions; in fact, the former solutions are thermodynamic excited states obtained from the latter by reversing $h$ with respect to $M$ \cite{purcel}. However, we remark that the $E$ {\it vs.} $T$ plot indicates that the zero-temperature behavior of the specific heat ($\frac{1}{2}$ per site) violates the third law of thermodynamics. 

We complete the description of the thermodynamics of the system by presenting the behavior of the Boltzmann entropy $S$ {\it vs.} $E$, for several values of $h$, derived using ~(\ref{self_consist})-(\ref{entropy}), and shown in figure 2(c). For $E_{\min}(h)\leq E \leq \frac{1}{2}^{-}$ ($\frac{1}{2}^{+}\leq E \leq \frac{1}{2} +h^{2}/2=E_{\max}(h)$) and $0\leq h\leq 1$, the solutions are stable for $T>0$ ($T<0$) with $\frac{\partial^2 S }{\partial E^2}<0$, while for $h\leq E \leq \frac{1}{2}+h^{2}/2$ the solutions are metastable (unstable) for $T>0$ with $\frac{\partial^2 S }{\partial E^2}<0$ \big($\frac{\partial^2 S}{\partial E^2}>0$\big); at the spinodal point $\frac{\partial^2 S }{\partial E^2}|_{T_{\max}}=0$. Here, the dash-dot line ($T=0$ line of critical points) corresponds to $S$ {\em vs.} $E_{\max}(h)$ for $0\leq h \leq 1$ and, similarly to figure 1(c), it is obtained by reflection of the $h=0$ Boltzmann entropy curve around $E_c =\frac{1}{2}$. As we shall show in the following section, the Euler entropy, $S_{\chi}(E;h)$, displays a similar pattern and allows the exact computation of the magnetic properties in the microcanonical ensemble.

\begin{figure*}[]
      \includegraphics[width=\linewidth,clip]{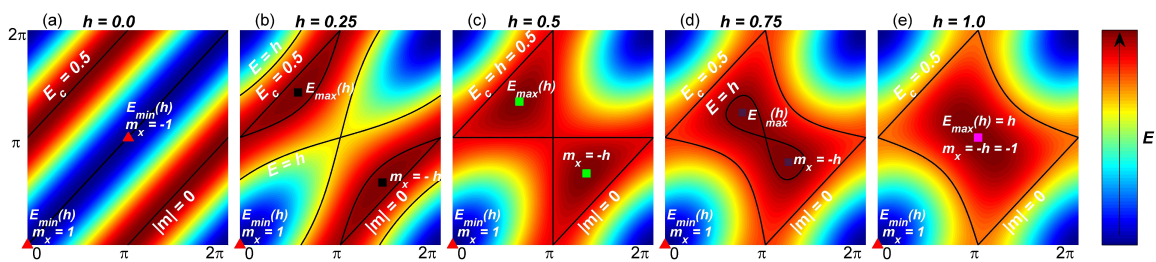}
      \caption{(Colour online) Infinite-range $XY$ model: level curves for $N=2$. The evolution of the level curves, which changes only at the critical points, are in correspondence with the thermodynamic solutions (see text).}
   \end{figure*}
\begin{figure}[t]
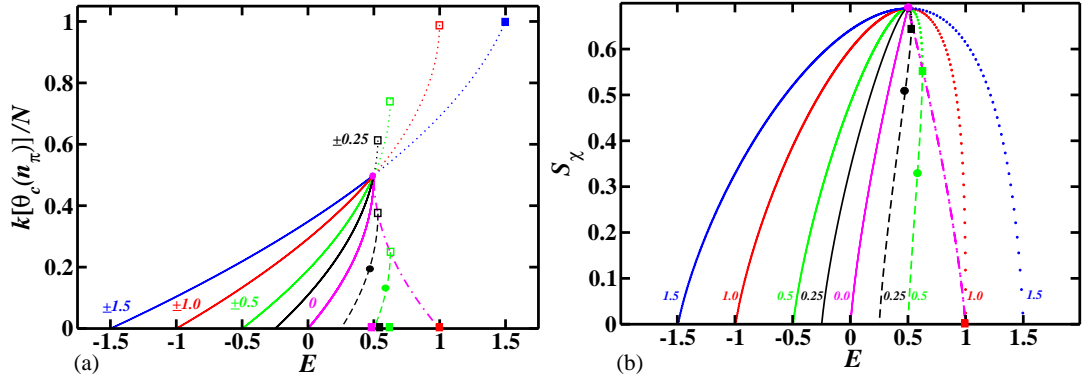

\centering
    \includegraphics[width=0.45\linewidth]{Fig4a.eps} \vspace{10 mm}
    \includegraphics[width=0.45\linewidth]{Fig4b.eps}
\caption{(Colour online) (a) Index $k[\theta_c(n_{\pi})]/N$ {\em vs.} $E[\theta_c(n_{\pi};h)]$.
 In the energy region corresponding only to stable (stable, unstable or metastable) solutions the diagram has only one branch (two branches) and $E(\theta; h)$ is (not) a {\em regular} Morse function. (b) Euler entropy $S_{\chi}(E)$ {\em vs.} $E$ [compare with the microcanonical  Boltzmann entropy in figure 2(c)]. In (a), $N=10^{3}$.}
\end{figure}
\begin{figure}
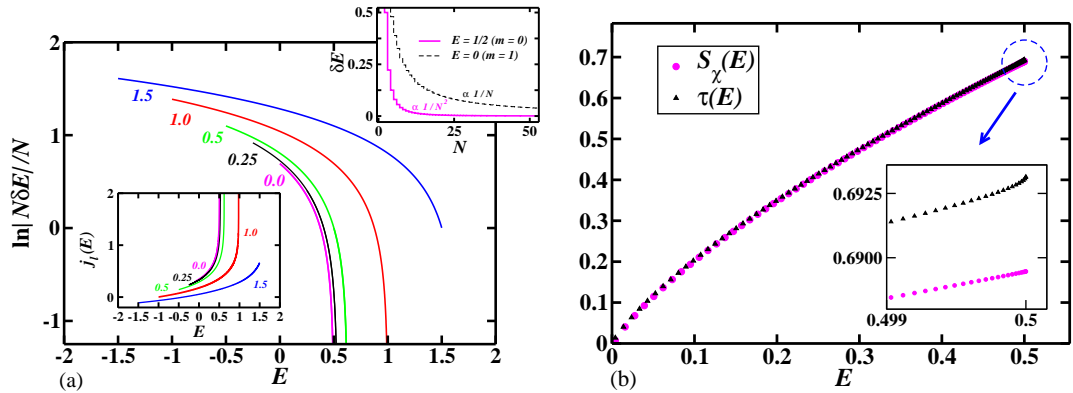

\centering
\includegraphics[width=0.45\linewidth]{Fig5a.eps}
    {\includegraphics[width=0.45\linewidth]{Fig5b.eps}}
\caption{(Colour online) (a) Divergence of $\ln|N\delta(E)|/N$ {\em vs.} $E$ and $j_{l}(E)$ {\em vs.} $E$ (lower inset) at $E = E_{\max}(h) = E_c+h^{2}/2$. In the upper inset we illustrate the change of scale in $\delta E$ at $E_c=\frac{1}{2}$ from $\mathcal{O}(\frac{1}{N})$ to $\mathcal{O}(\frac{1}{N^{2}})$. (b) Comparison between $S_{\chi}(E)$ and $\tau(E)$ {\em vs.} $E$. They are very close to each other (see inset); however, the value of $T_c=\frac{1}{2}$ is not obtained from the slope of $\tau(E\to E_c)$.}
\end{figure}

\section{Topological approach to the Microcanonical Thermodynamics and Phase Transition of the Infinite-Range $XY$ Model}
\subsection{The Case $N=2$}
Before studying the model in the thermodynamic limit, we find it instructive to present the topological
changes in the configuration space, $M_{E}$, of $E(\theta;h)$ in ~(\ref{energy_per_spin}) for its simplest case, i.e., $N=2$; thus $M_{E}$ can be seen as a surface in $\mathbb{R}^{3}$ and the Morse Theory (see Section II) promptly provides its evolution as a function of $E$.
\par For $N=2$, we have only four distinct $(\theta_{1_c},\theta_{2_c}$) critical configurations: $\{\mathbf{\theta_c}\}=\{(0,0),(0,\pi),(\pi,0),(\pi,\pi)\}$, plus the solution $\mathbf{m}=(m_x=-h$, $m_y=0$). According to the Noncritical Neck Theorem \cite{Palais}, the shape of $M_{E}$ will not change if there is no critical points in a given energy interval. Therefore, for each value of $h$ we should pay special attention to the shape of the critical equipotentials at the associated critical energy values, as illustrated for the equipotentials $M_E$ of $E(\theta;h)$ in figure $3$ for $h=0, 0.25, 0.5, 0.75$, and  $1.0$.
For $h=0$ all solutions are thermodynamically stable. In figure 3(a), we find that $M_{E}$ behaves as follows: if $E <0$, $M_E$ is an empty set; if $E(\theta_1=0,\theta_2=0;h=0)=E(\theta_1=\pi,\theta_2=\pi;h=0)=E_{\min}(h=0)=0$, we have two critical points (red triangles).
The nontrivial equipotentials for $h=0$ are straight lines. In fact, since $|M|=\sqrt{m_{x}^{2} + m_{y}^{2}}=\sqrt{|1-2E|}$, we can write $\theta_1=\theta_2 + \cos^{-1}(|1-4E|)$. The lines share a one-to-one correspondence between $|\mathbf{m}|$ and the energy level $E$. Further, if $E(\theta_1=\pi,\theta_2=0)=E(\theta_1=0,\theta_2=\pi)$, the surface reaches its highest level, $E_{\max}(h=0)=E_c=\frac{1}{2}$ (black straight lines)
and deserves highlights because these two lines are invariant even in the presence of $h$. In fact, the topological changes for $0< h\leq 1$ are illustrated in figures 3 (b) - 3 (e).
Notice that if $E <E_{\min}(h)=E(\theta_1= \theta_2=0;h)=-h$, $M_E$ is an empty set, while if $E=E_{\min}(h)$, we have a critical point at $m_x=1$ (red triangle). All equipotentials with energies in the range $E_{\min}(h) < E <E_c = \frac{1}{2}$ [$\frac{1}{2} \leq  E < E_{\max}(h)=\frac{1}{2}+h^{2}/2$] are simple (closed) curves;
 the black closed curves shown in figures 3(b)-3(e) are those associated with the special value $E_{c}=\frac{1}{2}$.
For $0<h<1$, the equipotential $E(\theta_1= \theta_2=\pi;h)=h$ is the unique non-simple curve, and crosses itself at the saddle point $m_x=-1$, as shown in figures 3 (b) - 3 (d); notice that for $E=h=\frac{1}{2}$ the referred curve degenerates with the curve $E_c=\frac{1}{2}$ [see figure 3(c)]. Moreover, equipotentials with energies in the range $E(\theta_1= \theta_2=\pi;h)=h\leq E \leq E_{\max}(h)$ are solutions corresponding to the van der Waals loops (metastable and unstable solutions for $T>0$) in the thermodynamic limit.
Last, the critical point $m_x =- h$, that appears as two maxima at $E_{\max}(h)=\frac{1}{2}+h^{2}/2$ in figures 3(b)-3(d), degenerates with the saddle point $m_x=-1$ at $E_{\max}(h)=h=1$ [figure 3(e)]. For $h\geq 1$, no relevant change in the configuration space is observed.

The results for our {\it smallest} interacting system show that, at $E_c=\frac{1}{2},$ a ``topological transition'' does take place and is characterized by the occurrence of closed curves for $E_c>\frac{1}{2}$. Further, the critical points that appear in figure 3(a)-3(c) are those found in figure 1(a)-1(c). Moreover, we emphasize that the equipotentials inside the black closed curves, limited by $E_c=\frac{1}{2}$, are high energy states corresponding to negative spin temperatures in the thermodynamic limit. Notice also that, since this system has only two neighboring spins, the MF solution is trivially identical to a XY 2-spin 1d system.

We find it instructive to calculate $\chi(M_{E,\delta E})$ and $\chi(\partial M_E)$ for $N=2$ in order to illustrate the procedure $\lim_{\delta E\to0}\chi(M_{E,\delta E})=\chi(\partial M_E)$. Notice that, while $\chi(M_{E,\delta E})$ is exactly calculated via Morse theory, the value of $\chi(\partial M_E)$ is calculated from the level curves in figure 3, via a triangulation technique. Besides the degenerate maxima at $E_{\max}=(\frac{1}{2}+h^2/2)$ [squares in figure 3(b)-(e)], there are the following three critical levels of $V(q)$ [assume, for simplicity, $h>0$]: (i) $E_{\min}=-h$, which is associated with the critical point $\theta_c(n_{\pi}=0)=(0,0)$ of index 0 [red triangles in figure 3(b)-(e)]: therefore $\chi(M_{E_{\min},\delta E})=1$. Here, $\partial M_E$ is a single point, and then $\chi(\partial M_{E_{\min}})=1$; (ii) $E_c=\frac{1}{2}$, which is associated with the two critical points $\theta_c(n_{\pi}=1)=(\pi,0)$ and $(0,\pi)$, both of index 1 [see the level curves at $E=E_c$ in figure 3(b)-(e)]: therefore, $\chi(M_{E_c,\delta E})=-2$. Here, $\partial M_{E_c}$ is a curve with $\chi(\partial M_{E_c})=-2$, since it can be triangulated with 2 vertexes, $(0,\pi)\equiv(2\pi,\pi)$ and $(\pi,0)\equiv(\pi,2\pi)$, and 4 edges connecting them; and (iii) $E_h=+h$, which is associated with the critical point $\theta_c(n_{\pi}=2)=(\pi,\pi)$ of index 1 [the saddles in figure 3(b)-(d) and magenta square in figure 3(e)]: and therefore $\chi(M_{E_h,\delta E})=-1$. Here, $\partial M_{E_h}$ is a ``figure 8 curve'' with $\chi(\partial M_{E_h})=-1$, since we can triangulate it with 1 vertex, $(\pi,\pi)$, and 2 (loop) edges. Notice that for $h>1$ the level curve $\partial M_{E_h}$ degenerates in a single point, while for $h=\frac{1}{2}$ the levels $E_c$ and $E_h$ coincide, figure 3(c). In this case, $\chi(M_{E_c,\delta E})=\chi(\partial M_{E_c})=-3$. In short, the limit procedure $\lim_{\delta E\to0}\chi(M_{E,\delta E})=\chi(\partial M_E)$ is verified and, despite that the critical level sets are odd-dimensional, $\dim\partial M_E=1$, the Euler characteristic does not vanish. We emphasize, however, that when $E$ is a regular value, the level set $\partial M_E$ is a closed $1\mbox{-manifold}$ and has a vanishing Euler characteristic, as can be verified in figure 3, in agreement with the prediction of the Poincar\'e Duality Theorem.



\subsection{Morse Theory and Topology of the Configuration Space}
We shall now use  Morse theory to describe the topology of the configuration space of the model at any finite $N$, and in the thermodynamic limit. The isolated critical points of  (\ref{first_derivate})
are $\theta=\theta_{c}=(\theta_{1_{c}},\ldots,\theta_{N_{c}})=\{0,\pi\}^{N}$, i.e., all
$\theta_{i}$ are either $0$ or $\pi$, plus the critical point $m_x \equiv M=-h$ under $m_y=0$.
Further, if we denote by $n_\pi$ the numbers of angles that are equal to $\pi$ in a given critical point,
the magnetization is written as 
\begin{equation}
 m_x \equiv M(n_{\pi}) =\big(1-\frac{2n_{\pi}}{N}),\label{def::magnetization}
\end{equation}
which implies that all magnetization values in the interval $[-1,1]$ are accessed for $N \gg 1$. Therefore, the potential energy per spin at the critical points of  (11) reads:
\begin{equation}\label{critical_energy}
E[\theta_{c}(n_{\pi});h]=\frac{1}{2}\big[1-(1-\frac{2n_{\pi}}{N})^{2}\big]-h(1-\frac{2n_{\pi}}{N}).
\end{equation}
On the other hand, for a fixed value of $E,$ we can invert the equation above in order to obtain the two possible solutions for $n_{\pi}$ \cite{CasettiJSP2003}:
\begin{equation}
 \bar{n}_{\pi}^{(\pm)}(E;h)=\frac{n_{\pi}^{(\pm)}(E;h)}{N}=\frac{1}{2}[1+h\pm\sqrt{h^2-2(E-\frac{1}{2})}]\,.\label{eq::def_npi_pm}
\end{equation}
Lets now examine the Morse number $\mu_{k}(M_E;h),$ which is the number of critical points lying in $M_{E}$ with $k$ negative eigenvalues of the Hessian:
$\mathcal{E}_{ij}=\frac{\partial^{2}E(\theta_c;h)}{\partial\theta_{i}\partial\theta_{j}}, \ i,j=1,...,N,$
 i.e., with index $k(\theta_c;h)$.
At a given critical point, and in the limit $N\gg 1$, the Hessian is
diagonal, with matrix elements given by \cite{CasettiJSP2003}:
$\mathcal{E}_{ii}[\theta_{c}(n_{\pi};h)]=\Big[\big(1-\frac{2n_{\pi}}{N}\big)+h\Big]\cos\theta_{i}$. Therefore,
\begin{equation}\label{index}
      k[\theta_c(n_{\pi};h)] =\left\{
      \begin{array}{ll}
      n_{\pi}^{(-)} & \mbox{if }n_{\pi}=n_{\pi}^{(-)} \\
      N-n_{\pi}^{(+)} &\mbox{if }n_{\pi}=n_{\pi}^{(+)} \\%
      \end{array}
      \right.;
   \end{equation}
   with multiplicity given by the Morse number \cite{CasettiJSP2003}:
\begin{equation}
   \mu_{k}(M_{E};h)={N\choose k}[1-\Theta(k-n_{\pi}^{(-)})+\Theta(N-k-n_{\pi}^{(+)})]\label{Morse_number}.
\end{equation}

The maximum value of $E$ is $E_{\max}(h)=\frac{1}{2}+h^2/2$ and, therefore, $\chi(M_E;h)=0$ since the configuration space is a $N\mbox{-torus}$; furthermore, it can be shown that \cite{CasettiJSP2003}, for $h\to0$,
\begin{equation}
\mu_k(M_E)=0 \mbox{ for all } k>N/2\,, 
\end{equation}
which implies that no critical points with index larger than $N/2$ exist as long as $E<E_c$. It then follows that $M_E$ is at most a ``half'' $N\mbox{-torus}$ for $\frac{1}{2}\leq E < \frac{1}{2}+h^2/2$ and a (full) $N\mbox{-torus}$ for $E=\frac{1}{2}+h^2/2$. We stress that the ``abrupt'' change at $E_{\max}(h)=\frac{1}{2}+h^2/2$ embodies the attachment of ${N\choose k}$ different $k\mbox{-handles}$ for each $k$ ranging from $N/2+1$ to $N$. This topology change at $E_c(h\to0)=\frac{1}{2}$ corresponds to the phase transition in the thermodynamic limit.

\par  Now, using  (\ref{eq::def_npi_pm}) and (\ref{index}), we can plot $k[\theta_c(n_{\pi};h)]/N$ vs. $E[\theta_c(n_{\pi};h)]$, as illustrated in figure 4(a)
for $h=0\,,0.25\,,0.5\,,1.0\,,1.5$. First, we notice that for $0\leq h\leq 1$ and $-h\leq E[\theta_c(n_{\pi};h)]<h$, i.e., $0\leq\frac{n_{\pi}}{N} <h$, $E(\theta;h)$ is a {\em regular} Morse
function, i.e., if $\theta_{1}$ and $\theta_{2}$ are critical points with $E(\theta_{1};h)<E(\theta_{2};h)$, we have $k(\theta_{1})<k(\theta_{2})$ \cite{Palais}. However, for $h\leq E[\theta_c(n_{\pi};h)] \leq \frac{1}{2}+h^{2}/2$, $E(\theta;h)$ is no longer a regular Morse function and we have two branches: the first associated with the $T>0$ ($T<0$) stable solutions for $n_{\pi}=n_{\pi}^{(-)} \in
\{0,\ldots,N/2\}$ ($n_{\pi}=n_{\pi}^{(-)} \in
\{N/2+1,\ldots,\lfloor\frac{  1 + h }{2}N\rfloor\}$), while the second branch corresponds to metastable and unstable solutions for $n_{\pi}=n_{\pi}^{(+)} \in \{\lceil\frac{1 +h}{2}N\rceil,\ldots,N\}$. Further, for $h> 1$ we again have only one branch corresponding to $T>0$ ($T<0$) stable solutions with $n_{\pi}=n_{\pi}^{(-)} \in \{0,\ldots,N/2\}$ ($n_{\pi}=n_{\pi}^{(-)} \in\{N/2+1,\ldots,N\}$). We remark that, by symmetry arguments, for $h<0$ the stable (metastable and unstable) solutions are associated with $n_{\pi}^{(+)}$ ($n_{\pi}^{(-)}$). The above results are in full qualitative agreement with the thermodynamic phase diagram shown in figure 2(a).  
\subsection{Euler Characteristic}
The separation of indexes into two distinct branches gives us hints on the subtle connection between the thermodynamic entropy and the Euler characteristic. In fact, by summing only over the critical points with indexes corresponding to each branch in figure 4(a), we split $\chi(M_E;h)$ into two distinct contributions: one corresponding to stable solutions and the other to metastable and unstable ones. 

Since the Euler characteristic of $M_E$ is a weighted sum over the Morse numbers,  (\ref{def::EulerChar}), it follows from  (\ref{Morse_number}) that
\begin{equation}
 \chi(M_E;h) = (-1)^{n_{\pi}^{(-)}}{N-1\choose n_{\pi}^{(-)}}+(-1)^{N-n_{\pi}^{(+)}}{N-1\choose N-n_{\pi}^{(+)}}\,,\label{eq::Chi_ME_h}
\end{equation}
where we have used the identities $\Theta(x-y)=1-\Theta(y-x)$ and $\sum_{k=0}^{m}(-1)^k{N\choose k} = (-1)^m{N-1\choose m}$. Now, observe that for $N\gg1$ we can approximate  (\ref{eq::Chi_ME_h}) by
\begin{equation}
 \chi(M_E;h)\cong (-1)^{n_{\pi}^{(-)}}{N\choose n_{\pi}^{(-)}}+(-1)^{N-n_{\pi}^{(+)}}{N\choose N-n_{\pi}^{(+)}}\label{eq::Chi_ME_h_FATIA}.
\end{equation}
It does corresponds to the Euler characteristic of $M_{E,\delta E},$  (\ref{def::EulerChar_Fatia}), which is the quantity we are interested in order to put forward its connection with the Boltzmann entropy. In fact, taking $\delta E>0$ smaller than the distance between neighboring critical points, $M_{E,\delta E}$ contains only those critical points with $n_{\pi}^{(-)}(E)$ and $n_{\pi}^{(+)}(E)$ angles equal to $\pi,$ i.e., with index $n_{\pi}^{(-)}$ and $N-n_{\pi}^{(+)}$, and multiplicity given by ${N\choose n_{\pi}^{(-)}}$ and ${N\choose N-n_{\pi}^{(+)}}$, respectively: 
\begin{equation}
 N\gg1\Rightarrow \chi(M_E;h)\cong\chi(M_{E,\delta E}\,;h).\label{eq::ApproxAsymp_Chi}
\end{equation}
Now we can safely take $\lim_{\delta E\to0}\chi(M_{E,\delta E})$ and associate the result of this limit with $\chi(\partial M_E)$.

In the thermodynamic limit, $S_{\chi}$ in  (\ref{equivBetweenEntropyANDSchi}) reads (suppose $h>0$):
\begin{eqnarray}
S_{\chi}(E;h) & = & \frac{1}{N}\ln\left\vert{N\choose n_{\pi}^{(-)}}\left(1+(-1)^{-Nh}\frac{{N\choose N-n_{\pi}^{(+)}}}{{N\choose n_{\pi}^{(-)}}}\right)\right\vert\nonumber\\
& = & \frac{1}{N}\ln{N\choose n_{\pi}^{(-)}}+\frac{1}{N}\ln\left\vert1+(-1)^{-Nh}\frac{{N\choose N-n_{\pi}^{(+)}}}{{N\choose n_{\pi}^{(-)}}}\right\vert\nonumber;
\end{eqnarray}
\begin{equation}
S_{\chi}^{(-)}(E;h>0)=-\bar{n}_{\pi}^{(-)}\ln\bar{n}_{\pi}^{(-)}
 -(1-\bar{n}_{\pi}^{(-)})\ln(1-\bar{n}_{\pi}^{(-)}),\label{eq::S_chi_ThermoLimit}
\end{equation}
where $\bar{n}_{\pi}^{(\pm)}(E;h)$ is defined in  (\ref{eq::def_npi_pm}) and, to obtain  (\ref{eq::S_chi_ThermoLimit}), we have used that $h>0$ implies ${N\choose N-n_{\pi}^{(+)}}/{N\choose n_{\pi}^{(-)}}\to0$, for $N\to\infty$ \cite{LimitBinomial}. This limiting process corresponds to a Maxwell construction to the van der Waals loops. In fact, we stress that the second term in  (\ref{eq::Chi_ME_h_FATIA}) is associated with metastable and unstable solutions, and should be calculated separately:
\begin{equation}
S_{\chi}^{(+)}(E;h>0) = -\bar{n}_{\pi}^{(+)}\ln\bar{n}_{\pi}^{(+)}
 -(1-\bar{n}_{\pi}^{(+)})\ln(1-\bar{n}_{\pi}^{(+)}).\label{eq::S_chi_ThermoLimitMeta}
\end{equation}
Notice that the expressions for both $S_{\chi}^{(-)}$ and $S_{\chi}^{(+)}$ are formally identical to that of a Fermi system with $0\leq \bar{n}_{\pi}^{(\mp)} \leq1$ \cite{Kubo}, and positive definite. A similar expression also appears in the context of the k-trigonometric model \cite{AngelaniPRE2005,LuizkTM}. The origin of this fact is due to Morse theory and the occurrence of isolated critical points $\theta=(\theta_{1},...,\theta_{N})$ of $V(\theta;h)$ with only two values for $\theta_i$: $0$ or $\pi$.   

For $h<0$ the stable (metastable and unstable) solutions are associated with $n_{\pi}^{(+)}$ ($n_{\pi}^{(-)}$). In zero field, the two contributions to the Euler characteristic in  (\ref{eq::Chi_ME_h}) or (\ref{eq::Chi_ME_h_FATIA}) are identical, since $n_{\pi}^{(-)}(E;h\to0)=N-n_{\pi}^{(+)}(E;h\to0)$. Therefore, in the thermodynamic limit, $S_{\chi}(E;h\to0)$ can be obtained from either  (\ref{eq::S_chi_ThermoLimit}) or  (\ref{eq::S_chi_ThermoLimitMeta}). We remark that, since at $E=E_{\max}$, $M_{E_{\max}}$ is a $N$-Torus, $\chi(M_{E_{\max}})=0$. However, for practical purposes, we extend $\chi(M_{E};h)$ at $E=E_{\max}$ by continuity of  (\ref{eq::Chi_ME_h}); likewise for $S_\chi$ and other quantities defined in terms of $\chi(M_E;h)$. In figure 4(b), we plot $S_{\chi}(E;h)$ which is the analog of the microcanonical entropy, in figure 2(c), in the context of our topological approach, including stable, unstable and metastable solutions. Indeed, these solutions occur in the {\em same} energy intervals and with the {\em same} pertinent convexities [compare figures 2(c) and 4(b)]. These results strongly suggest that the {\em loss of regularity} in the Morse function (potential energy) is associated with the occurrence of unstable and metastable solutions. We remark that the dashed-dot magenta line in figure 4(a) or 4(b) corresponds to $k[\theta_c(n_{\pi};h)]$ or $S_{\chi}(E;h)$ {\em vs.} $E_{\max}(h)$ for $0\leq h\leq 1,$ respectively; similarly to the entropy plot in figure 2(c), this line is also obtained by reflection of the zero field line around $E_c=\frac{1}{2}$. 

Notice that, differently from the singular behavior of the Boltzmann entropy, the Euler entropy (and the corresponding specific heat) vanishes as $E\to E_{\min}$ by construction; notwithstanding, both the Boltzmann and the Euler temperatures vanish for $E\to E_{\min}$. In this context, we remark, without further interpretation, that the thermodynamic weight is calculated using the continuous energy spectrum, while the Euler characteristic is calculated using the discrete energy levels associated with the critical values of the potential. Additionally, $S_{\chi}(E;h)\to\ln2$ as $E\to E_c$ due to discrete nature of the critical energy levels, while, as expected, $S(E;h)\to\ln(2\pi)$, figures 4(b) and 2(c), respectively.

Now we study the quantity $\tau(E;h)$ defined in  (\ref{def::approxEntByNc}). We shall focus on the calculation of the stable solutions associated with $n_{\pi}^{(-)}$; therefore, 
\begin{equation}
 N_c(E;h,N) = \sum_{j=0}^N \mu_j(M_E;h) = \sum_{j=0}^{n_{\pi}^{(-)}}{N\choose j}.\label{eq::Nc}
\end{equation}
Then, we have, for $h>0$ and $n_{\pi}^{(-)}(E;h)\leq N/2$:
\begin{equation}
{N\choose n_{\pi}^{(-)}} \leq  N_c(E;h)
\end{equation}
and
\begin{equation}
N_{c}(E;h) = {N\choose n_{\pi}^{(-)}}\left\{1+\sum_{j=0}^{n_{\pi}^{(-)}-1}\underbrace{{N\choose j}\Big/{N\choose n_{\pi}^{(-)}}}_{\leq1}\right\}
 \leq {N\choose n_{\pi}^{(-)}}\{1+n_{\pi}^{(-)}\}\label{eq::StepAux_ApAsym_Tau}.
\end{equation}
However, since $0\leq n_{\pi}^{(-)}\leq N$, 
\begin{equation}
 \lim_{N\to\infty}\frac{1}{N}\ln(1+n_{\pi}^{(-)})=0,
\end{equation}
and we thus find that for $N\gg1\mbox{ and }n_{\pi}^{(-)} \leq N/2$:
\begin{equation}
\tau(E;h) = \frac{1}{N}\ln N_{c}(E;h) \sim \frac{1}{N}\ln{N\choose n_{\pi}^{(-)}(E;h)}\label{eq::ApproxAsymp_Tau}.
\end{equation}
Therefore, in the thermodynamic limit:
\begin{equation}
 \tau(E;h)=-\bar{n}_{\pi}^{(-)}\ln \bar{n}_{\pi}^{(-)}-(1-\bar{n}_{\pi}^{(-)})\ln(1-\bar{n}_{\pi}^{(-)}),
\end{equation}
which is {\em identical} to $S_{\chi}(E;h)$ in  (\ref{eq::S_chi_ThermoLimit}), under the condition $\bar{n}_{\pi}^{(-)}(E;h) \leq \frac{1}{2}$. For $h<0$ the stable (metastable and unstable) solutions are associated with $n_{\pi}^{(+)}$ ($n_{\pi}^{(-)}$).

On the other hand, for $N\gg1$ and $\bar{n}_{\pi}^{(-)}(E;h)>\frac{1}{2},$
\begin{equation}
 \tau(E;h) = \frac{1}{N}\ln N_{c}(E;h) \sim \ln2.\label{eq::tauSaturates}
\end{equation}
Therefore, in the thermodynamic limit, $\tau(E;h)$ saturates for $E\geq\frac{1}{2},$ in contrast to the predictions from both the Boltzmann entropy,  (\ref{entropy}) and figure 2(c), and the Euler entropy $S_{\chi}(E;h)$,  (\ref{eq::S_chi_ThermoLimit}) and figure 4(b).

\par Last, we find it instructive to study the energy difference between two arbitrary neighboring critical levels, i.e.,
   \begin{equation}
   \label{delta}
    \Delta E = E[\theta_c(n_{\pi});h]-E[\theta_c(n_{\pi}-1);h] = \frac{2}{N}[M(n_{\pi})+h]+\frac{2}{N^{2}}.
   \end{equation}
As illustrated in figure 5(a), $\ln|N\Delta E|/N$ {\em vs.} $E$ diverges for points along the line of critical points $M=-h$, i.e., $E_{\max}(h)$ for $0\leq h\leq 1$. Notice that for $M(n_{\pi}) = -h$, $\Delta E$ is of $\mathcal{O}(\frac{1}{N^{2}})$, including at the PT:
$M(n_{\pi})=0$ and $h=0$, as shown in the upper inset of figure 5(a) for $E_c=\frac{1}{2}$.
 Further, we can also study the saddle-point contributions from the critical points in the neighborhood of $E$ to the entropy, i.e., the density of Jacobian's critical points, $j_{l}(E)$, given by \cite{KastnerPRL2007}
   \begin{equation}\label{jacobian_density}
      j_{l}(E)
      =\lim_{N\to\infty}\frac{1}{N}\ln\biggl(\sum\limits_{\theta_c \in
      Q_l(E,E+\epsilon)}\!J(\theta_c)\;\bigg/
      \!\!\sum\limits_{\theta_{c}\in Q_l(E,E+\epsilon)}1\biggr),
   \end{equation}
where $J(\theta_c)$ is the Jacobian determinant, $Q_l(E,E+\epsilon)$
denotes the set of critical points $\theta_c$ with index $k(\theta_c)=l\pmod 4$
and with critical values $E(\theta_c)/N$ in the interval
$[E,E+\epsilon]$. For the present model,
$j_{l}(E)$ can be derived analytically \cite{KastnerPRL2007}:
$j_{l}(E)=\frac{1}{4}\ln\Big| \frac{2}{E-(\frac{1}{2}+h^2/2)}\Big|.$ 
As shown in the lower inset of figure 5(a), $j_{l}(E)$ indeed diverges for $0\leq h \leq 1$ at $E_{\max}(h)=\frac{1}{2}+h^{2}/2$ \cite{SantosPRE2009}.
 Noticeably, $\ln|N\Delta E|/N$ and $j_{l}(E)$ diverge at the same critical values.
   In fact, we find that
  \begin{equation}
  \ln|N\Delta E|/N = 2[\ln2-j_{l}(E)].
\end{equation}

\subsection{Microcanonical Magnetization via Euler Measure\label{sec::chiMag}}

 \par With the aim to assess thermodynamic quantities, such as the magnetization, from a topological viewpoint, we can compute the mean magnetization, $M_{\chi}(E;h)$, using either an average over $M_E$ or $M_{E,\delta E}$,  (\ref{vm}), in agreement with  (\ref{eq::ApproxAsymp_Chi}). In fact, averaging over $M_E$, we have
 \begin{equation}
  M_{\chi}(E;h) = \sum_{\{\theta_c:E(\theta_c;h)\leq E\}}  \frac{(-1)^{k(\theta_c;h)}\Big[1-\frac{2n_{\pi}(\theta_c;h)}{N}\Big]}{\chi(M_E;h)}.\label{eq::Mchi_ME}
 \end{equation}
Taking into account the index $k[\theta_c(n_{\pi};h)]$,  (\ref{index}), and the degeneracy of the critical points given by the Morse numbers,  (\ref{Morse_number}), we can write 
\begin{eqnarray}\label{M_chi}
M_{\chi}(E;h)  
 & = & \frac{1}{\chi}\Big\{\sum_{k=0}^{n_{\pi}^{(-)}}(-1)^k\big(1-\frac{2k}{N}\big){N\choose k}+
 \sum_{k=n_{\pi}^{(+)}}^{N}(-1)^{N-k}\big(1-\frac{2k}{N}\big){N\choose k}\Big\}\nonumber\\
& = & \frac{(-1)^{n_{\pi}^{(-)}}{N-1\choose n_{\pi}^{(-)}}+2(-1)^{n_{\pi}^{(-)}-1}{N-2\choose n_{\pi}^{(-)}-1}}{(-1)^{n_{\pi}^{(-)}}{N-1\choose n_{\pi}^{(-)}}\left(1+(-1)^{-Nh}\frac{{N-1\choose N-n_{\pi}^{(+)}}}{{N-1\choose n_{\pi}^{(-)}}}\right)}-\nonumber\\
& - & \frac{(-1)^{N-n_{\pi}^{(+)}}{N-1\choose N-n_{\pi}^{(+)}}+2(-1)^{N-n_{\pi}^{(+)}-1}{N-2\choose N-n_{\pi}^{(+)}-1}}{(-1)^{N-n_{\pi}^{(+)}}{N-1\choose N-n_{\pi}^{(+)}}\left(1+(-1)^{Nh}\frac{{N-1\choose n_{\pi}^{(-)}}}{{N-1\choose N-n_{\pi}^{(+)}}}\right)},\label{eq::Aux_MagChi}
\end{eqnarray} 
where we have made the substitution $(N-k)\to k$ in the second sum of the first equality and used the identities: $\sum_{k=0}^{m}(-1)^k{N\choose k} = (-1)^m{N-1\choose m}$ and $\sum_{k=0}^{m}(-1)^kk{N\choose k}=N(-1)^m{N-2\choose m-1}.$ 

Now, in analogy with the procedure that leads to  (\ref{eq::ApproxAsymp_Chi}), we find for $N\gg1$
\begin{eqnarray}
M_{\chi}(E;h) & \cong & \left(1+(-1)^{-Nh}\frac{{N\choose N-n_{\pi}^{(+)}}}{{N\choose n_{\pi}^{(-)}}}\right)^{-1}\left\{1-2\frac{\bar{n}_{\pi}^{(-)}}{1-1/N}\right\}-\nonumber\\
&-&
\left(1+(-1)^{Nh}\frac{{N\choose n_{\pi}^{(-)}}}{{N\choose N-n_{\pi}^{(+)}}}\right)^{-1}\left\{1-2\frac{1-\bar{n}_{\pi}^{(+)}}{1-1/N}\right\}\\\label{for::MagChiIntegral_Field}
& = & \frac{1}{\chi(M_{E,\delta E};h)}\sum_{\vert E(\theta_c;h)-E\vert\leq\delta E}(-1)^{k(\theta_c;h)}\Big[1-\frac{2n_{\pi}(\theta_c;h)}{N}\Big].\label{eq::Aux2_MagChi}
\end{eqnarray}
Since $h>0$ implies $\lim_{N\to\infty}{N\choose N-n_{\pi}^{(+)}}/{N\choose n_{\pi}^{(-)}}=0$ \cite{LimitBinomial}, in the thermodynamic limit, the stable solution reads:
\begin{equation}
 M_{\chi}^{(-)}(E;h>0) = \Big(1-2\bar{n}_{\pi}^{(-)}(E;h)\Big),\label{eq::Mchi_m}
\end{equation}
where $\bar{n}_{\pi}^{(\pm)}(E;h)$ is defined in  (\ref{eq::def_npi_pm}). This corresponds to the Maxwell construction solution to the thermodynamic magnetization. On the other hand, similarly to the analysis for $k[\theta_c(n_{\pi};h)]$ and $S_{\chi}(E;h)$, the metastable and unstable solutions are obtained from the second term of  (\ref{eq::Aux2_MagChi}), and should be calculated separately:
\begin{eqnarray}
 M_{\chi}^{(+)}(E;h>0) 
 & = & \Big(1-2\bar{n}_{\pi}^{(+)}(E;h)\Big).\label{eq::Mchi_p}
\end{eqnarray}
On the other hand, for $h<0$ the stable (metastable and unstable) solutions are associated with $n_{\pi}^{(+)}$ ($n_{\pi}^{(-)}$).

Remarkably, the expressions for $M_{\chi}(E;h)$, whether stable, metastable, or unstable solutions are identical to the Boltzmann microcanonical thermodynamic ones in  (\ref{eq::Mthermo_pm}).

In analogy with $M_{\chi}(E;h)$, we can also define a mean magnetization via the entropy $\tau(E;h)$ [$M_{\tau}(E;h)$]. In fact, the expression for it reads
\begin{equation}
 M_{\tau}(E;h) = \frac{1}{N_c(E;h,N)}\sum_{\{\theta_c:E(\theta_c;h)\leq E\}}(1-\frac{2n_{\pi}(\theta_c)}{N}),
\end{equation}
where $N_c(E;h,N)$ is given in  (\ref{eq::Nc}). We shall focus on the calculation of the stable solutions associated with $n_{\pi}^{(-)}$, i.e. $h>0$.

Taking into account the index $k[\theta_c(n_{\pi};h)]$,  (\ref{index}), and the degeneracy of the critical points given by the Morse numbers,  (\ref{Morse_number}), we can write
\begin{eqnarray}
  M_{\tau}(E;h) 
  & = & \frac{1}{N_c(E;h,N)}\sum_{k=0}^{n_{\pi}^{(-)}}\left(1-\frac{2k}{N}\right)\mu_k(M_E)\nonumber\\
  & = & 1-\frac{1}{N_c(E;h,N)}\frac{2}{N}\sum_{k=0}^{n_{\pi}^{(-)}}k\,{N\choose k}\nonumber\\
  & = & 1-\frac{2}{N_c(E;h,N)}\sum_{k=0}^{n_{\pi}^{(-)}-1}{N-1\choose k}\nonumber\\
  & = & 1-2\frac{N_c(n_{\pi}^{(-)}(E)-1;h,N-1)}{N_c(n_{\pi}^{(-)}(E);h,N)},
 \end{eqnarray}
where we have used the identity $k{N\choose k}=N{N-1\choose k-1}$, and the substitution $(k-1)\to k$, in the second equality. 

For large $N$, it is possible to estimate the value of $M_{\tau}$ in order to compare it with $M_{\chi}$. Assuming $h>0$ and $n_{\pi}^{(-)}\leq N/2$, the value of ${N\choose n_{\pi}^{(-)}}$ [ ${N-1\choose n_{\pi}^{(-)}}$ ] is much greater than the values of ${N\choose k}$ [ ${N-1\choose k}$ ] when $k\leq n_{\pi}^{(-)}$ [ $n_{\pi}^{(-)}-1$ ] and, therefore, one has
 \begin{eqnarray}
  M_{\tau}(E;h) \approx1-2\frac{{N-1\choose n_{\pi}^{(-)}-1}}{{N\choose n_{\pi}^{(-)}}}=1-2\frac{n_{\pi}^{(-)}(E;h)}{N}.
 \end{eqnarray}
On the other hand, if $n_{\pi}^{(-)}>N/2$, then it follows that
\begin{eqnarray}
 M_{\tau}(E;h) \approx 1-2\frac{{N-1\choose N/2-1}}{{N\choose N/2}} = 1-2\frac{N/2}{N} = 0.
\end{eqnarray}

In summary,
\begin{equation}
 M_{\tau}(E;h) =\left\{\begin{array}{cc}
                        1-2\bar{n}_{\pi}^{(-)} = M_{\chi} &\mbox{ if }\,\,\bar{n}_{\pi}^{(-)}\leq \frac{1}{2}\\
                        0 & \mbox{ if }\,\,\bar{n}_{\pi}^{(-)}> \frac{1}{2}\\
                       \end{array}
                       \right..
\end{equation}
Thus, $M_{\chi}$ coincides with $M_{\tau}$, derived using $\tau(E;h)$ instead of $\chi(M_E;h)$, only for $E\leq\frac{1}{2}$. In addition, $M_{\tau}$ vanishes precisely in the region where $\tau(E;h)$ saturates, which is consistent with the fact that $1/T_{\tau}=\partial \tau(E;h)/\partial E=0$, i.e. $T_{\tau}=\infty$, for $E>\frac{1}{2}$. This result is in agreement with the approach using $\Omega_0(E)$ and the proposal in  \cite{DunkelNaturePhys2014} to exclude the occurrence of states with negative spin temperature, but in contradiction with the prediction using the Boltzmann entropy.  

\subsection{Critical Temperature}

In closing our analysis of the model, we calculate the PT critical temperature of the system using the Euler characteristic. For this purpose, we compute the Euler temperature, $T_{\chi}$, as a function of the energy $E$. In order to capture the smallest topology change in the computation of $T_{\chi}$ and, therefore, to identify a topological analogous for an infinitesimal change, we must study the change induced by those critical points with the same index. Indeed, the topology change at $E=E_{\max}$ embodies the attachment of different kinds of indexes, as emphasized in the discussion following  (\ref{Morse_number}). Therefore, at $h=0$ and near $E_{c}=\frac{1}{2}$, $T_{\chi}$ can be computed analytically using the handle decomposition of the configuration space \cite{Matsumoto} and unveils a direct relationship between the Euler characteristic and the PT critical temperature. First, we notice that $\chi(M_{E_{c}=\frac{1}{2}^{+}})=\chi(M_{E_{c}=\frac{1}{2}})+(-1)^{k=N/2}\mu_{k=N/2}(M_{E_{c}=\frac{1}{2}})=0$ [see figure 4(b)], where we have not considered the contribution of those critical points with index greater than $N/2$. Without loss of generality, we now use ~(\ref{Morse_number}) for $N=2n$ and $n_{\pi}=N/2=n$, to obtain
  $\chi(M_{E_c=\frac{1}{2}})=-(-1)^n{2n\choose n}$.
Similarly, for the closest critical energy level below $E_c=\frac{1}{2}$, i.e., $E_{c}-\Delta E_c$, associated with the Morse numbers $n_{\pi}=N/2\pm
1$ [see ~(\ref{critical_energy}) for $h=0$], we have,
\begin{equation}
\chi(M_{E_{c}})=\chi(M_{E_{c}}-\Delta E_c)+\sum_{k\in\{N/2\pm1\}}(-1)^{k}\mu_{k}(M_{E_{c}-\Delta E_c}),
\end{equation}
which implies
   $|\chi(M_{E_c-\Delta E_c})|=2{2n\choose n-1}-{2n\choose n}$.
Last, by noticing that $\Delta (E_c) = E(n_{\pi}=n)-E(n_{\pi}=n \pm 1) = 1/2n^{2}$ [see ~(\ref{critical_energy})], the Euler critical temperature reads:
\begin{eqnarray}
\frac{1}{T_{\chi_c}} 
   & = &\lim_{n\rightarrow\infty}\frac{1}{2n}\frac{\ln\frac{|\chi(M_{E_c})|}{|\chi(M_{E_c-\Delta E})|}}{\Delta E}=\lim_{n\to\infty}\frac{1}{2n}\frac{\ln\left(\frac{{2n\choose n}}{2{2n\choose n-1}-{2n\choose n}}\right)}{\frac{1}{2n^2}}\nonumber\\
& = & \lim_{n\to\infty}n\ln\left(2\frac{{2n\choose n-1}}{{2n\choose n}}-1\right)^{-1}=\lim_{n\to\infty}\ln\left(\frac{1+1/n}{1-1/n}\right)^n,
\end{eqnarray}
which, by using  $e^{x}=\lim_{n\rightarrow\infty}
\Big(1+\frac{x}{n}\Big)^{n}$,
we finally obtain
\begin{equation}\label{tc}
T_{\chi_{c}}=T_{c}=\frac{1}{2}.
\end{equation}
\par We remark that the Euler weight, $(-1)^{k}$, was a crucial feature in order to prove that the critical temperature $T_c=T_{\chi(E_c)}$ from the topology of $M_ {E}$. In fact, as $\tau(E)=\ln2^N$ at $E=E_c^+=\frac{1}{2}$ and $\tau(E_{c}=\frac{1}{2}^{+})=\tau(E_{c}=\frac{1}{2})+\mu_{k=N/2}(M_{\frac{1}{2}})$, if we attempt to find $T_c$ from  (\ref{def::approxEntByNc}): 
\begin{eqnarray}
\frac{1}{T_{\tau_c}} 
    &=& \lim_{n\rightarrow\infty}\frac{\tau(E_c)-\tau(E_c-\Delta E)}{\Delta E}\nonumber\\
&=& \lim_{n\to\infty}\ln\left(\frac{2^{2n}-{2n\choose n}}{2^{2n}-{2n\choose n}-2{2n\choose n-1}}\right)^n\nonumber\\
&=&\lim_{n\to\infty}\ln\left(\frac{1-\frac{{2n\choose n}}{2^{2n}}}{1-\frac{3n+1}{n+1}\frac{{2n\choose n}}{2^{2n}}}\right)^n,
\end{eqnarray}
we obtain
\begin{equation}\label{tauc}
T_{\tau_{c}}=0\not=T_{c}=\frac{1}{2}.
\end{equation}
Then, the behavior of the corresponding $T_{\tau}$ fails just near $E_{c}$, as can be inferred from the comparison of the behavior of $\tau(E)$ and that of $S_{\chi}$ {\em vs.} $E$ in figure 5~(b). We stress that the difference between $\tau(E)$ and $S_{\chi}(E)$ is negligible except for $E$ very close to $E_{c}$, in which case the slope of $\tau(E)$ differs significantly from that of $S_{\chi}(E)$, i.e., $\lim_{E\to E_c}\tau'(E)=\infty$, as illustrated in the inset of figure 5(b); notice also that, from  (\ref{eq::tauSaturates}), $\tau'(E>E_c;h>0)=0$ (infinite temperature), while, from  (\ref{eq::S_chi_ThermoLimit}), $S_{\chi}'(E>E_c;h>0)<0$ (negative temperature).

\section{Topological Approach to the Microcanonical Thermodynamics and Phase Transition of the $1d$ short-range $XY$ model}

We shall now use the topological approach presented above to describe the 1d classical planar short-range $XY$ model \cite{StanleyXY1d}, whose potential energy reads ($J\equiv1/4$):

\begin{equation}\label{eq:H_XY_1d}
V(\theta;h)=\frac{1}{4}\sum_{i=1}^N [1-\cos(\theta_{i+1}-\theta_i)] -h\sum_{i=1}^N\cos(\theta_i).
\end{equation}
In the thermodynamic limit, the zero field microcanonical entropy per spin is given by \cite{Mehta}
\begin{equation}
S(E)=\frac{\beta(E)}{4}(4E-1)+\ln\Big[2\pi I_0(\beta(E)/4)\Big],
\end{equation}
where $E$ is the energy per spin and $\beta(E)$ is the solution of the self-consistency equation:
\begin{equation}\label{1d_self_consist}
(1-4E)=\frac{I_1}{I_0}(\beta(E)/4).
\end{equation}
Notice that the thermodynamic entropy exhibits a singular behavior as $E\to0$, while $S(E)\to \ln2\pi$ as $E\to 0.25$. Moreover, only at sufficiently low-$T$ does the system obeys the Equipartition Energy Theorem, with the specific heat in agreement with the MF result. 

\par The analysis of the topology of the configuration space and the critical points of the model were reported in  \cite{CasettiJSP2003} and \cite{Mehta}, respectively. In calculating the critical points of  (\ref{eq:H_XY_1d}) we must exploit the rotation invariance of the model. In fact, in  \cite{Mehta} the symmetry was broken by fixing $\theta_N = 0$, while in \cite{CasettiJSP2003} a small field $h\rightarrow 0 $ was applied. We choose the latter approach since we find it more suitable in computing $M_{\chi}$. The critical points of $V(\theta;h)$ in  (\ref{eq:H_XY_1d}) are, as done for the infinite-range $XY$ model, $\theta_c=\{0,\pi\}^{N}$. However, the critical energy values are determined by the number of boundaries between regions with the same orientation, i.e., the number $n_d$ of domain walls. 

It is important to mention that, at $h = 0$, in addition to the solutions $\{0, \pi\}^N$, one can find other families of critical points. In fact, the total number of critical points is \cite{Mehta}:
\begin{equation}
\#(\theta^s)=\sum_{j=0}^{N-1}\vert N-2j \vert {N-1\choose j}=\frac{N!}{\{[(N-1)/2]!\}^2}\,.
\end{equation}
Now, by noticing that the total number of  $0 - \pi$ solutions is $2^{N}$, the difference between the logarithm per site of the total number of critical points and that of $0 - \pi$ ones is:
\begin{equation}
\frac{1}{N}\ln \#(\theta^s) -\frac{1}{N}\ln 2^{N} \sim \frac{\ln N}{N},
\end{equation}
which goes to zero in the thermodynamic limit. Additionally, the full range of energy is accessed for $N \gg 1$ by the critical values of the $0 - \pi$ solutions. In conclusion, these results suggest that the investigation of the $0- \pi$ solutions suffices to determine the thermodynamic properties of the model using the Euler Measure. In fact, as will become clear in the following, this is indeed a suitable choice.

From now on, we consider the family $\{0, \pi\}^N$ as the only pertinent set of critical solutions. The energy per particle in a given critical point reads:
\begin{equation}\label{eq:End}
E[\theta_c(n_d,n_{\pi});h]=\frac{n_d}{2N}-h\Big(1-\frac{2n_{\pi}}{N}\Big).
\end{equation} 
In addition, the number of critical points with $n_d$ domains is $2{N-1\choose n_d}$, and the index $k$ of a point in configuration space with $n_d$ domain walls is $k=n_d$, for $h\to0$; then, the Morse numbers and the Euler characteristic are \cite{CasettiJSP2003} 
\begin{equation}
\mu_k(M_E)=2{N-1\choose k}\Theta(n_d(E)-k)\,.
\end{equation} 
and:
\begin{equation}\label{eq:chi}
\chi(M_{E})=\sum_{k=0}^N(-1)^{k}\mu_k(M_E)=2(-1)^{n_d(E)}{N-2\choose n_d(E)},
\end{equation}
respectively; where, for further use, the Zeeman term is considered in  (\ref{eq:End}).

\subsection{Mean Magnetization}

In zero field, $E[\theta_c(n_d,n_{\pi});h=0]=n_d/2N$ and we can thus find the topological invariants directly as function of the energy $E$. Therefore, since $m(n_{\pi})=(1-2n_{\pi}/N)$,
in computing $M_{\chi}$ using
 (\ref{eq:chi}), we have
\begin{equation}\label{eq:M_chi_XY_1d}
M_{\chi}(E)=\chi(M_E)^{-1}\sum_{n_d\leq \lfloor 2NE\rfloor}\!\!\!\!\!\!\!\!(-1)^{n_d}\Big[\!\!\sum_{n_{\pi}\in \mathcal{D}}\!\!\Big(1-\frac{2n_{\pi}}{N}\Big)W(n_d,n_{\pi})\Big],
\end{equation}
where $\mathcal{D}(n_{d},n_{\pi}) \ \ [W(n_{d},n_{\pi})]$ is the set (multiplicity) of $n_{\pi}$'s configurations with a fixed $n_d$.
In fact, $W(n_{d},n_{\pi})$ can be obtained from  \cite{RehnBJP2012,DenisovPRE1993} and is given by
\begin{equation}
W(n_{d},n_{\pi}) = \left\{
\begin{array}{cc} 
{n_{\pi}-1\choose n_d/2-1}{N-n_{\pi}-1\choose n_d/2}+{n_{\pi}-1\choose  n_d/2}{N-n_{\pi}-1 \choose n_d/2-1}, & \mbox{ if } n_{d} \mbox{ is even} \\[6pt] 2{n_{\pi}-1\choose (n_d-1)/2}{N-n_{\pi}-1\choose (n_d+1)/2}, & \mbox{if } n_{d}\mbox{ is odd} 
\end{array}
\right.,
\end{equation}
which for a fixed value of $n_d$ satisfies 
\begin{equation}
\sum_{n_{\pi}}W(n_d,n_{\pi})=2{N-1\choose n_d}.
\end{equation}
As expected, summing over $n_{\pi}$ and $n_d$, one obtains
\begin{equation}
 \sum_{n_{\pi},n_d}W(n_d,n_{\pi})=\sum_{n_d}2{N-1\choose n_d}=2^N.
\end{equation}

The inversion of all spins by the transformation $n_{\pi}\rightarrow N-n_{\pi}$ maps $m(n_{\pi})$ on $-m(n_{\pi})$ and preserves the number of domain walls, i.e., for a fixed $n_d$, we have the same number of configurations with opposite magnetizations. In conclusion, $M_ {\chi}$ in  (\ref{eq:M_chi_XY_1d}) is zero for $ E \neq 0$.
However, the case $E=0$ ($n_d=0$), which corresponds either to $\theta_i=0, \forall i$ [$m(n_{\pi}=0)=1$] or to $\theta_i=\pi, \forall i$ [$m(n_{\pi}=N)=-1$], must be carefully analyzed.
In the presence of a small field  $h\rightarrow 0$, we have $E(\theta_i=0)=0-h$ and $E(\theta_i=\pi)=0+h$, with an energy gap  $2h$ between the $m=\pm 1$ configurations; $h$ should be smaller than the distance between critical levels, i.e., $h<1/2N$, in order to avoid overlapping between critical levels. Due to the Noncritical Neck Theorem \cite{Palais}, the topological invariants for $-h < E <h $ are distinct from those with ($E>h$). So, to compute $M_{\chi}(E)$ for the first critical point, we have to choose an energy value in the interval $(-h,h)$, which, due to the fact that we have only one critical point below the chosen energy level, it thus turns out that $M_{\chi}$ is equal to $1$. In short,
\begin{equation}
M_{\chi}(E)=\left\{
 \begin{array}{ccc}
  1 &,& E=0\\[4pt]
  0 &,& E>0\\
 \end{array}
\right..
\end{equation}

\begin{figure}[t]
\centering
\includegraphics[width=0.6\linewidth,clip]{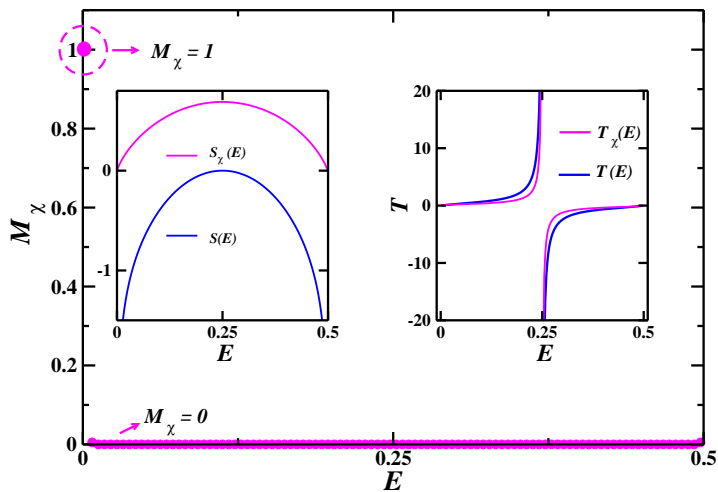}
\caption{(Colour online) Short-range planar $XY$ model. $M_{\chi}$ {\em vs.} $E$ diagram, displaying the expected phase transition at $T=0$. In the first inset we plot the Euler and Boltzmann entropies per site, $S_{\chi}(E)$ and $S(E)$ (in fact, $S(E)-\ln2\pi$), respectively, {\it vs.} energy per site $E$; in the magnified area around $E=0$, we show the data of $S_{\chi}(E)$ for $N=10^{5}$, indicating an almost vertical slope, i.e., $T_{\chi}\to0$ as $E\to0$. In the second inset we show the plot of $T$ and $T_\chi$ {\em vs.} $E$.}
\end{figure}
\begin{figure}[t]
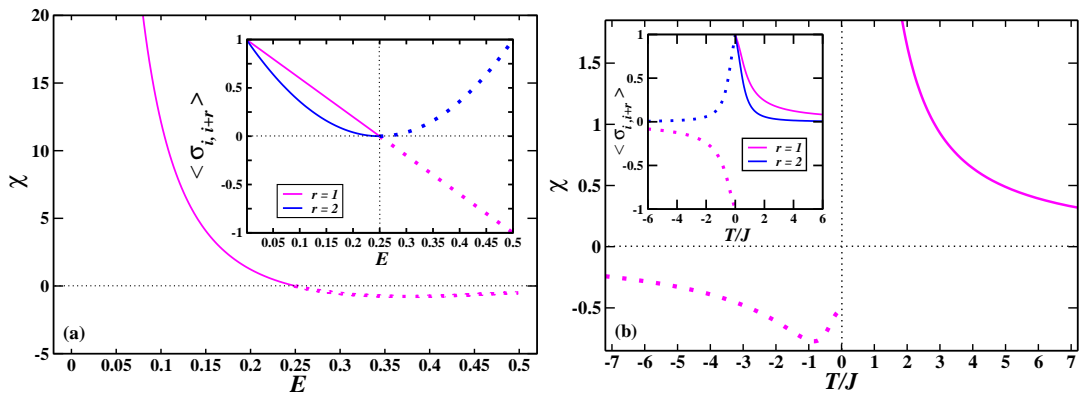

\centering
{\includegraphics[width=0.45\linewidth,clip]{Fig7a.eps}}
{\includegraphics[width=0.45\linewidth,clip]{Fig7b.eps}}
\caption{(Colour online). Short-range planar XY model: (a) microcanonical susceptibility, $\chi\,vs.\,E$. (b) canonical susceptibility, $\chi\,vs.\,T$. The insets present the corresponding correlation functions, $\langle\sigma_{i,i+r}\rangle,$ for $r=1$ and $r=2$.}
\label{susceptibility}
\end{figure}

The above argument does not hold for $n_{d}\neq 0$. Since we now have $2{N-1\choose n_d}$ critical points, the small field $h$ does not affect the natural distribution of critical points with opposite magnetizations. In fact, for $n_d=0$ we have only two critical points, while for $n_d\neq 0$ the number of critical points grows asymptotically with $N$ as $N^{n_d}$ and the field effect is thus negligible in computing $M_{\chi}$.

Here we point out that the calculation of $M_{\tau}(E)$ is similar to that of $M_{\chi}(E)$, i.e., one just removes the factor $(-1)^{n_d}$ in  (\ref{eq:M_chi_XY_1d}), and verify that this procedure does not change the result of the magnetization: $M_{\tau}(E) = M_{\chi}(E)$. 

The diagram $M_{\chi}$ (or $M_{\tau}$) {\em vs.} $E$ is illustrated in figure 6 for zero field. In the first inset, we compare the microcanonical entropy and the Euler characteristic. It is clear that a phase transition takes place at $T=0$. 

\subsection{Euler Temperature}

Now, we apply the definition of the Euler temperature,  (\ref{eulert}), to the 1d short-range $XY$ model. Then, using  (\ref{eq:chi}),  we have
\begin{equation}
\frac{1}{T_{\chi}(E)}=\lim_{N\rightarrow \infty}\frac{1}{N}\frac{\ln|\chi(n_d +1)|-\ln|\chi(n_d)|}{\Delta E},
\end{equation}
where $\Delta E =1/2N$, which follows from  (\ref{eq:End}) in zero field. We thus find that
\begin{equation}
T_{\chi}(E)=\lim_{N\rightarrow \infty}\frac{1}{2\ln\Big(\frac{N-2-n_d}{n_d+1}\Big)}.
\end{equation}
Since for $h=0$ we have $E=n_d/2N$, in the thermodynamic limit $T_{\chi}$ as a function of the energy reads:
\begin{equation}\label{Tchi}
T_{\chi}(E)=\frac{1}{2\ln\big(\frac{1}{2E}-1\big)};
\end{equation}
thereby 
\begin{equation}
\displaystyle\lim_{E\rightarrow 0}T_{\chi}(E)=T_{c}=0.\end{equation}
Likewise, we can similarly calculate $T_{\tau}$ at $E=0$:
\begin{equation}
T_{\tau}(E=0)=0=T_{\chi}(E=0)\,.
\end{equation}

As in the MF case, but differently from the singular behavior of the Boltzmann entropy in the thermodynamic limit, for the 1d short-range $XY$ model the Euler entropy vanishes as $E$ approaches zero:
\begin{equation}
 S_{\chi}(E) = -\bar{n}_d\ln \bar{n}_d-(1-\bar{n}_d)\ln(1-\bar{n}_d)\stackrel{E\to0}{\longrightarrow}0,\label{eq::chi1d}
\end{equation}
where $\bar{n}_d(E)=\frac{n_d(E)}{N}=2E$, and use was made of  (\ref{eq:chi}). Notice the similarity of  (\ref{eq::chi1d}) with the MF expressions for $S_{\chi}$,   (\ref{eq::S_chi_ThermoLimit}) and (\ref{eq::S_chi_ThermoLimitMeta}). 
In addition, we can also calculate $\tau(E)$, which reads:
\begin{equation}
\tau(E)=\left\{
 \begin{array}{ccc}
  S_{\chi}(E) &,& E\leq\frac{1}{4}\\[4pt]
  \ln2 &,& E\geq\frac{1}{4}\\
 \end{array}
\right..
\end{equation}
We emphasize that in the 1d case, the phase transition occurs at $T_c=0$, i.e. $E_c=0$, while the saturation of $\tau(E)$ occurs at the energy $E=0.25$, in which case both the Boltzmann and Euler entropies attain  their maximum value: $\ln 2\pi$ and $\ln 2$, respectively.
This feature is common to both MF and short-range versions.
Lastly, from  (\ref{eq::chi1d}) , we obtain
\begin{equation}
\frac{\partial S_{\chi}(E)}{\partial E} = -2\ln(2E) +2\ln(1-2E).
\end{equation}
Therefore, $\lim_{ E \rightarrow 0^{+}} \frac{\partial S_\chi (E)}{\partial E}\sim  -2\ln2 E \stackrel{E\to0^+}{\longrightarrow}  \infty$ and  $\lim_{ E \rightarrow \frac{1}{2}^{-}} \frac{\partial S_\chi (E)}{\partial E}\sim  2\ln(1-2 E) \stackrel{E\to{\frac{1}{2}}^{-}}{\longrightarrow} -\infty$, in agreement with the $T$-range discussed in section II.B, which implies, in the microcanonical ensemble, that $S_{\chi}(E)$ approaches zero at the transition critical point, $E_c = T_c = 0$, trough either of the two limits just described. We thus concluded that, by joining the points $E_c=0$ ($T_c=0^+$) and $E=\frac{1}{2}$ ($T=0^-$), equivalent to the translation: $S_{\chi} \equiv S_{\chi}(E-\frac{1}{2})$ for $E \in [\frac{1}{4},\frac{1}{2}) $ and $S_{\chi}(E) \equiv S_{\chi}(E)$ otherwise, $S_{\chi}$(E) exhibits an umbilical point with a singular cusp behavior \cite{LivroShape} in the vicinity of the transition critical point. This feature holds true also for the Bolztmann entropy, regardless of its singular behavior at $T=0$, as numerically verified by the $T, T_{\chi}$ vs. $E$ diagram for both entropies (see the second inset of figure 6). In addition, one should stress that the MF {\it finite $T_c$} occurs at $T_{\max }$ of the Euler and microcanonical Bolztmann entropies, and cusp around $T_c$ is characterized by a {\it finite} discontinuity of the entropy derivatives (See figures 2c and 4b). On the other hand, while the determinant $D$ of the Hessian is zero at the MF finite {$T_c$} (see the  inset of figure 5(a), $D=1$ for the 1-d short range $XY$ model PT at the $T_c =0$. In any case, the topological hypothesis is preserved (see discussion in section I).

\subsection{Correlation Function and Susceptibility}
We want to study the following pair correlation function:
\begin{equation}
\langle\sigma_{i,i+1}\rangle = \frac{1}{N} \sum_{i=1}^{N}\cos(\theta_i-\theta_{i+1}),
\end{equation}
evaluated at the critical points $\{0,\pi\}^{N}$. We see that, for a given configuration, $\langle\sigma_{i,i+1}\rangle$ can be written in terms of the number of domain walls $n_d$ due to the following properties: i) for pairs of spins which are displayed outside the domain walls, we have $\cos(\theta_i-\theta_{i+1})=1$; ii) for pairs of spins displayed between domain walls, i.e., with opposite spins, we have $\cos(\theta_i-\theta_{i+1})=-1$. Now, notice that we have $(N-n_d)$ spins in situation i) and $n_d$ spins in situation ii); therefore, we get:
\begin{equation}
 \langle\sigma_{i,i+1}\rangle = 
 \frac{1}{N}\Big[(N-n_d)(1)+n_{d} (-1)\Big] = \Big( 1- \frac{2 n_{d}}{N} \Big).
\end{equation}
The above result, put together with  $(\ref{eq:chi})$, allow us to compute the correlation function analytically through the Euler measure,  (\ref{vm}):
\begin{equation}
 \langle\sigma_{i,i+1}\rangle_{\chi}(E)=\frac{1}{\chi(M_E)}\!\!\!\sum_{n_d\leq \lfloor NE/2\rfloor}\!\!\!(-1)^{n_d}\Big(1-\frac{2 n_d}{N}\Big)\mu_{n_d}(M_E).\label{eq::CorrelationChi}
\end{equation}
Now, using the identity $\sum_{k=0}^{m}(-1)^k{N\choose k} = (-1)^m{N-1\choose m},$ we find that
\begin{equation}
 \langle\sigma_{i,i+1}\rangle_{\chi}(E)\!\!=\!\!\frac{\Big[(-1)^{n_d(E)}{N-2\choose n_d(E)}+2 (-1)^{n_d(E)-1} {N-3\choose n_d(E)-1}  \Big]}{(-1)^{n_d(E)}{N-2\choose n_d(E)}}.
\end{equation}
Following steps similar to those in Section \ref{sec::chiMag}, we thus get in the thermodynamic limit:
\begin{equation}
 \lim_{N\rightarrow \infty} \langle\sigma_{i,i+1}\rangle_{\chi}(E) = \lim_{N\to\infty}\Big[1 - 2 \bar{n}_{d}(E)\Big]=1-4E,\label{eq::1stCorrelation}
\end{equation}
where $\bar{n}_d(E)=n_d(E)/N.$ We also remark that in zero field and periodic boundary conditions the energy per spin of the Ising model, which can be mapped on a two level system, is given by $-1+n_d/2N$ (for $J=1$) \cite{RehnBJP2012}. 

Comparing  (\ref{1d_self_consist}) and (\ref{eq::1stCorrelation}), we have for the pair correlation function via the $\chi$-integral:
\begin{equation}
 \langle\sigma_{i,i+1}\rangle_{\chi}(E)=1-4E = \frac{I_{1}}{I_{0}}(\beta/4),\label{eq::1stCorrelationChiBeta}
\end{equation}
which is the well known result for the correlation function of the 1d XY model \cite{StanleyXY1d}. Moreover, since the system is 1d and the interaction is among first neighbors only, $\langle\sigma_{i,i+r}\rangle = \langle\sigma_{i,i+1}\rangle^{r}$, and the canonical susceptibility reads \cite{StanleyXY1d}: 
\begin{equation}
 \chi(\beta) = \frac{\beta}{2}\left[\frac{1+y(\beta/4)}{1-y(\beta/4)}\right],
\end{equation}
where $y(\beta/4)=\frac{I_1}{I_0}(\beta/4),$ while the microcanonical version is obtained with the help of  (\ref{1d_self_consist}).

The susceptibility is plotted in figures \ref{susceptibility}(a) and 7(b) in the microcanonical and canonical ensemble, respectively, while the correlation functions are shown in the corresponding insets for $r=1$ and $r=2$. As expected, for $T>0,$ the ferromagnetic susceptibility diverges as $\chi(E)\cong 1/2E^2+O(1/E)$ [$\chi(T)\cong 2/T^2+O(1/T)$], thus signaling the phase transition at $T=0$, in agreement with Fisher's results for the classical Heisenberg model \cite{FisherAJP1964}. On the other hand, for $T<0$, the antiferromagnetic susceptibility displays a maximum value $\chi_{\max}=0.7730$ at $T_{\max}/J=0.8348$, in agreement with Stanley's exact solution for a linear chain of interacting classical spins of arbitrary dimensionality  \cite{StanleyXY1d} [see e.g. figure 5(a), and normalization factors used, leading to a $\chi_{\max}$ and $T_{\max}$ twice bigger than ours]. Our $\chi_{\max}$ value for the classical $XY$ model should be compared with that found for the classical Heisenberg model 
\cite{FisherAJP1964,StanleyXY1d}: $\chi_{\max}=1.2045$ and $T_{\max}/J=0.2382.$ 

At this point it is instructive to notice that, from a formal point of view, the first-neighboring correlation function,  (\ref{eq::CorrelationChi}), is similar to the MF magnetization,  (\ref{eq::Mchi_ME}) and (\ref{eq::Mchi_MEnplus}), if one replaces $n_d$ by $k$ and vice-versa; however, they are not similar as a function of $E$, since $n_d$ {\it vs.} $E$ is not equal to $k$ {\it vs.} $E$. Therefore, we can use this similarity to conclude that:
\begin{equation}
 \langle \sigma_{i,i+1}\rangle_{\tau}(E)=\left\{
 \begin{array}{ccc}
  \langle \sigma_{i,i+1}\rangle_{\chi}(E) &,& E\leq\frac{1}{4}\\[4pt]
  0 &,& E\geq\frac{1}{4}
 \end{array}
\right..\label{eq::1stCorrelationTau}
\end{equation}
In addition, since $\tau(E)$ imposes an infinite temperature in the region $E\geq0.25$, it implies a zero value for the correlation function $\langle\sigma_{i,i+1}\rangle_{\tau}$ and the corresponding susceptibility as well.

In short, the results for the magnetization, critical temperature correlation function and susceptibility in zero field for the 1d short-range $XY$ model show complete compatibility between the Boltzmann thermodynamic description and the topological approach using the Euler entropy. In particular, the spin states in the negative temperature region are related to those in the microcanonical ensemble with energies greater than $E=0.25$, via  (\ref{1d_self_consist}), and are indeed formally mapped onto the corresponding antiferromagnetic states at positive temperature. We emphasize that, in the microcanonical ensemble, the PT displayed in figure 6 is characterized by $M(E) = 0$ for $E = n_d /2N > 0$ and a sudden change to the saturated value of the magnetization at $E=0$: $M(E=0) = 1$; the spin-rotational  invariance is broken in the presence of  $h\rightarrow 0$, as expected for the classical 1d XY  model \cite{StanleyXY1d}. The first inset in figure 6 shows that $S_{\chi}(E)$ is zero at $E=0$, attains its maximum value at $S_{\chi}(E=0.25) = \ln 2$,  and decreases to zero at $E=0.5$; correspondingly, in this energy interval the determinant of the Hessian $D(E)$ is  defined by two symmetrical straight lines with respect to $E= 0$: from $D(E=0) =1$ to $D(E=0. 25) = 0$, and from $D(E=0. 25) = 0$ to $D(E= 0.5) =1$ (see figure 2 in  \cite{Mehta}). In addition, as shown in the second inset, the plot $T$ vs. $E$ exhibits two branches with singular behavior at $E = 0.25$, corresponding to the crossing of the spin temperature from $+\infty$ to $-\infty$. Therefore, $D$ is zero only at this singular value of energy.  The singular behavior at $E = 0.25$ is also manifested in figure 7 through the change of sign of the microcanonical susceptibility and first-neighbor spin correlation  function (see the inset of figure 7).  A complementary view of the phenomena around the $T=0$ PT is provided via the canonical ensemble, as shown in figure 7: a divergent ferromagnetic susceptibility on the $T > 0$ side, and a characteristic AF susceptibility on the $ T < 0$ side.

\section{Discussion and Conclusions}

We have proposed a topological approach which have allowed us to establish connections between thermodynamics and the topology of configuration space in the microcanonical ensemble. In the last sections, we have reported and discussed on results that point to the possibility of describing the statistical mechanics of interacting classical spin systems in the thermodynamic limit, including the occurrence of a phase transition, using topology arguments only. Our approach relies on Morse theory, through the determination of the critical points of the potential energy, which is the proper Morse function. Our main finding is to show that, in the context of the studied exactly solvable classical models, the Euler entropy $S_{\chi}(E;h)$, defined by the logarithm of the modulus of the Euler characteristic per site, exhibits subtle connections with the Boltzmann microcanonical entropy, and allows the exact computation of the magnetic properties, such as the magnetization, susceptibility, correlation function, and critical temperature, using the Euler measure. Further, in the MF case, the results suggest that the loss of regularity in the Morse function is associated with the occurrence of unstable and metastable thermodynamic solutions. 

The reliability of our approach was tested in two classical systems exactly soluble by standard methods of statistical mechanics: (i) the infinite-range and (ii) the short-range $XY$ models in the presence of a magnetic field. In (i), we have showed the equivalence of the microcanonical and canonical ensembles, including metastable, unstable state, and negative spin temperature states. Remarkably, our topological approach was shown to be consistent with the thermodynamic description using the Boltzmann entropy. Indeed, in the thermodynamic limit, in contrast to the entropy $\tau(E;h)$ calculated as a sum of the Morse numbers, which saturates for energies above the critical one, the predictions from both the Boltzmann microcanonical entropy and the Euler entropy are dominated by the multiplicity of the microscopy states at a {\em fixed} value of energy. Therefore, unlike the description using the Boltzmann and Euler entropies, which allows for negative spin temperature states, the use of $\tau(E;h)$ replaces these states by one of infinite temperature in the entire energy region where $\tau(E;h)$ saturates, similarly to the behavior of $\Omega_0(E)$ in the region of high-energy states. Moreover, the correct value for the thermodynamic critical temperature of the model was also found by using the Euler entropy. However, we stress that the results derived using $S_{\chi}(E;h)$ and $\tau(E;h)$ are in full agreement for $E<E_c$, and that all the subtleties discussed above are due to the classical nature of the models (continuous energy spectrum), the critical point being a special one in the case of zero field. In (ii), within the same scope, our proposed topological approach was suitable to treat a classical interacting $XY$ spin chain exhibiting a zero temperature phase transition, including the spin correlation function and magnetic susceptibility. In particular, we remark that the spin states in the negative temperature region are related to the high-energy microcanonical ones and are indeed formally mapped onto the corresponding antiferromagnetic states at positive temperature, similarly to the 1d Ising model \cite{RehnBJP2012}. In contrast to the MF case, the correct thermodynamic magnetization and critical temperature are obtained from both Euler and $\tau(E)$ entropies. This is so because, for the short-range XY model, $T_c=0$. On the other hand, the approach using $\tau(E)$ imposes an infinite temperature in the region $E\geq0.25$, which gives rise to a zero value for the correlation function and the corresponding susceptibility as well. 


In conclusion, the above results strongly suggest that, for quantities 
that do not violate the laws of thermodynamics, such as those related to the {\it magnetic} properties of the system, our proposal based on a purely Euler topological approach proved to be an alternative to describe the microcanonical thermodynamics and phase transition of {\it interacting classical spin systems}. Moreover, We confirm that the topological hypothesis \cite{Pettinilivro} holds for both the infinite-range ($T_c \neq 0$) and the short-range ($T_c = 0$) XY models. In fact, despite that the Boltzmann entropy is singular at $T=0$, while the Euler entropy is zero, both entropies give rise to identical magnetic properties. Other quantities, such as the specific heat in the studied $XY$ models, are strongly affected by the classical nature of the systems. On the other hand, it is rewarding that for the 1d Ising model \cite{RehnBJP2012}, which exhibits a discrete symmetry, the referred entropies, and consequently the correspondent low-T specific heat behavior, are identical in the thermodynamic limit.
In addition, besides the models analyzed in the present work, very recently we have also verified that the referred approach describes the k-trigonometric model in the same fashion as it did for the above-mentioned $XY$ models \cite{LuizkTM}.  Finally, pertinent generalizations of the approach and studies of other classical or semi-classical interacting systems are very desirable. These studies will certainly shed light on the validity of the proposed topological approach beyond the referred studied models, and on the necessity of possible generalizations. 

\appendix
\setcounter{section}{1}

The results presented on the integration with respect to the Euler characteristic in this work are analytic and sound. Despite that there are other applications of Euler integration \cite{Baryshnikov,bell}, to the best of our knowledge, this is the first application of this integration in the context of the topological approach to thermodynamics and phase transitions.  In order to achieve a proper understanding of this integration, it will prove instructive to apply this concepts to simple two-dimensional compact surfaces of genus $g$, previously studied  in two distinct situations \cite{CasettiJSP2003}: i) when the attachment  of handles occurs uniformly as we cross the critical levels, in analogy with the $1d - XY$ model; and ii) when we have an attachment of a high number of handles at the same critical level, in analogy with the $MF-XY$ model. We start with the Torus, i.e.,  $g=1$, when the function to be integrated is constant; further, we also investigate two cases with an arbitrary genus $g$. These topological two-dimensional models are natural choice to pedagogically illustrate the Euler integration, since they were first introduced  as a means to get insights on the topological approach to phase transitions.

\subsection{Example on the Torus $\mathbb{T}^2$}
\label{sec:examples}

\begin{figure}[ht]
\centering
\includegraphics[width=0.4\textwidth]{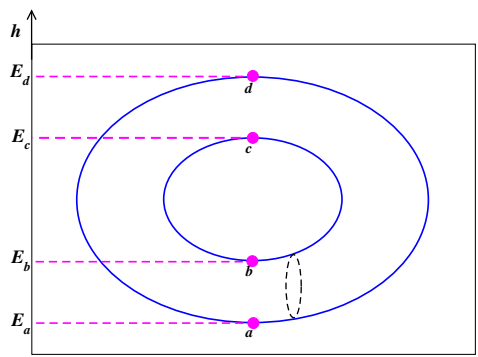}
\caption{\label{fig:T2} The height function $h$ on the torus $\mathbb{T}^2$.}
\end{figure}

We start by analyzing  the simple case of a Torus $\mathbb{T}^2$ and the its height function $h$ as the Morse function, which has four critical points: a minimum, a maximum, and two saddles, with critical values $E_a<E_b<E_c<E_d$ (figure A1). Here, we follow the standard procedure to describe the Torus using tools from Morse theory \cite{Matsumoto}, together with concepts of Euler Integral \cite{Livro}. Due to the Noncritical Neck Theorem, if there is no critical point on a given interval, no topology change occurs. That said, given a function $f$, the calculation of $\int_{\mathbb{T}^2}\,f{\rm d}\chi$ only depends on the existence of critical points of the (height) function $h$ and we can split the evaluation of $\int_{\mathbb{T}^2}\,f{\rm d}\chi$ in four steps: when $E\in[E_a,E_b)$, $E\in[E_b,E_c)$, $E\in[E_c,E_d)$, and $E\in[E_d,\infty)$, which corresponds to the nontrivial critical intervals (see figure A1). From now on, we denote as the critical set of the height function $h$ the set $\{(q, k(q))\}$, where $k(q)$ is the index of a critical point $q$ of $h$. 

Let $f\equiv1$ be a constant function on the torus. For $E\in[E_a,E_b)$,  the critical  set is $\{(a,0)\}$. We thus get
\begin{equation}
\int_{M_E}1\,{\rm d}\chi =\int_{M_E}\,{\rm d}\chi= \sum_{(q,k)}(-1)^{k(q)}\cdot 1 = 1.
\end{equation}
For $E\in[E_b,E_c)$, the critical set is $\{(a,0),(b,1)\}$:
\begin{equation}
\int_{M_E}\,{\rm d}\chi = \sum_{(q,k)}(-1)^{k(q)}\cdot 1 = (-1)^0+(-1)^1=0.
\end{equation}
Note that, contrary to the Riemann integral, the condition $f>0$ does not imply $\int_M f\,{\rm d}\chi>0$ when $f$ is continuous. In fact, it is even possible to have $f>0$ while $\int_M\,f\,{\rm d}\chi<0$. Indeed, in our example, for $E\in[E_c,E_d)$, the critical set is $\{(a,0),(b,1),(c,1)\}$, while
\begin{equation}
\int_{M_E}\,{\rm d}\chi = \sum_{(q,k)}(-1)^{k(q)}\cdot 1 = 
                     -1<0.
\end{equation}
Finally, for $E\in[E_d,\infty)$, the critical set is $\{(a,0),(b,1),(c,1),(d,2)\}$:
\begin{equation}
\int_{M_E}\,{\rm d}\chi = \sum_{(q,k)}(-1)^{k(q)}\cdot 1 =
\chi(\mathbb{T}^2)=0.
\end{equation}
Again, the integral vanishes even though the integrating function is positive. These results show that the comparison between the Euler and Riemann integral may be subtle.


\subsection{Examples on genus $g$ surfaces}
\label{sec:genusg}

\begin{figure}[t]
\centering
\includegraphics[width=0.45\textwidth]{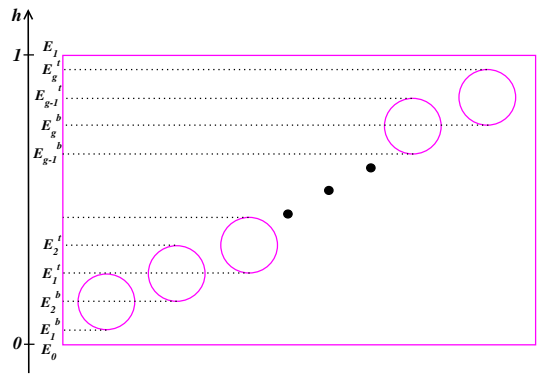}
\caption{\label{fig:T2g}The height function $h$ on a ``regular'' genus $g$ closed surface $\mathbb{T}_g^2$.}
\end{figure}

Let us now analyze  the genus $g$ surfaces used in figures 9 and 11 of  \cite{CasettiJSP2003} to illustrate the potential effect of attaching a significant number of handles on the behavior of the Euler characteristic in comparison with a situation in which 
only a small number of handles are attached. These
surfaces were chosen as toy models for the 1d short-range and infinite-range $XY$ models, respectively. Here we will study the dependence  of the Euler Integral on $g$ and the limit $g\gg1$ as well.

The first surface of interest, $\mathbb{T}_g^2$,  is a compact surface with $g$ holes uniformly distributed, as illustrated in figure A2, while the other surface, $\mathbb{T}_{g,\epsilon}^2$, is a deformation of the first one, i.e., the top of each hole is deformed in such a way that the critical points corresponding to them have the same height $h_{\max}-\epsilon$, as illustrated in figure A4 (for simplicity  we take $h_{\max}=1$). Note that the distance between two neighboring levels in $\mathbb{T}_g^2$ (see figure A2) is always the same, say $\Delta E$, and except for the two initial levels $E_1^b$ and $E_2^b$, each new hole adds one more level to the height of the surface (here $E_i^b$ and $E_i^t$ stand for the levels of the bottom and the top of the $i\mbox{-th}$ hole, respectively, as illustrated in figure A2). Then, we can distributed the levels as 
\begin{equation}
E_0<E_1^b<E_2^b<E_1^t<E_2^t<...<E_k^t<...<E_g^t<E_1,\label{eq::distLevels}
\end{equation}
which implies a total number of $g+4$ levels. It follows that 
\begin{equation}
\Delta E=\frac{1}{3+g}\stackrel{g\gg1}{\to}0.
\end{equation}
Therefore,
\begin{equation}
E_i^t=\frac{i+2}{3+g};\,E_i^b=E_i^t- \displaystyle\frac{2}{3+g}=\frac{i}{3+g}
,\, i=1,\cdots,g.
\end{equation}
Notice that those critical points corresponding to $E_i^b$ and $E_i^t$ have index 1, while those corresponding to $E_0=0$ and $E_1=1$ have index $0$ and $2$, respectively.
Consequently, since we are studying two-dimensional surfaces, this toy model does not illustrate the role played by the attachment of saddles with high order indexes. Additionally, we have the relation $E_i^t=E_{i+2}^b$ for $1\leq i\leq g-2$. 

\subsubsection{The constant function $f\equiv1$ on $\mathbb{T}_g^2$.}
Let $f\equiv1$ be the constant $1$ function on $\mathbb{T}_g^2$. Noticing that the levels can be arranged according to  (\ref{eq::distLevels}), we calculate $\int f\,{\rm d}\chi$ for $E\in[E_0,E_1^b)$, $E\in[E_1^b,E_2^b)$, $E\in[E_2^b,E_1^t)$, $E\in[E_i^t,E_{i+1}^t)$ with $i\in\{1,...,g-1\}$, $E\in[E_g^t,E_1)$, and $E\in[E_1,\infty)$.

For $E\in[E_0,E_1^b)$, one has only the level $E_0$ corresponding to a point with index 0:
\begin{equation}
\chi(M_E) = \int_{M_E}\,{\rm d}\chi=(-1)^0=1.
\end{equation}
For $E\in[E_1^b,E_2^b)$, one has also $h(q_1^b)=E_1^b$ of index $1$:
\begin{equation}
\chi(M_E) = \int_{M_E}\,{\rm d}\chi=(-1)^0+(-1)^1=0.
\end{equation}
For $E\in[E_2^b,E_1^t)$, one has also $h(q_2^b)=E_2^b$ of index $1$:
\begin{equation}
\chi(M_E) = \int_{M_E}\,{\rm d}\chi=(-1)^0+2(-1)^1=-1.
\end{equation}
\begin{figure}[t]
\centering
\includegraphics[width=0.45\textwidth]{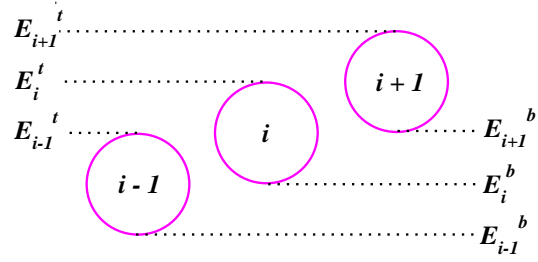}
\caption{\label{fig:ithHole}The situation around a generic hole of $\mathbb{T}_g^2$.}
\end{figure}

Now we perform the calculations for $E\in[E_i^t,E_{i+1}^t)$, where $i\in\{1,...,g-1\}$. At a critical point $q_i^t$, with $h(q_i^t)=E_i^t$, we have the following levels  below it: $E_0$, $E_1^t,\cdots,E_{i}^t$ and $E_1^b,\cdots,E_{i+1}^b$; if $i<g-1$, we also have $E_{i+2}^b$ (see figure A3). So, for $E\in[E_i^t,E_{i+1}^t)$, we have
\begin{equation}
\chi(M_E) = \int_{M_E}\,{\rm d}\chi=\left\{
\begin{array}{ccc}
-1-2i & , & i<g-1\\
+2-2g   & , & i=g-1\\
\end{array}
\right..
\end{equation}
Further, for $E\in[E_g^t,E_1)$,  the levels $E_0,E_1^t,...,E_g^t$ and $E_1^b,...,E_g^b$ are below the level $E_g^t$. Then,
\begin{equation}
\chi(M_E)=\int_{M_E}\,{\rm d}\chi=(-1)^0+g(-1)^1+g(-1)^1=1-2g.
\end{equation}
Finally, for $E\geq E_1$, we find
\begin{equation}
\chi(M_E)=\int_{M_E}{\rm d}\chi=(1-2g)+(-1)^2=2-2g,
\end{equation}
which is the well known result for the Euler characteristic of an orientable closed surface of genus $g$.

In summary, the values of $\chi(M_E)$ can be  arranged in the table below:
\begin{equation}
\begin{array}{|c|c|c|c|c|c|c|c|c|c|}
\hline
i\mbox{-th}\Delta E & 0 & 1 &2&3&...&g&g+1&g+2&g+3\\
\hline
\chi(M_E) &1&0&-1&-3&\,...\,&-3-2g&2-2g&1-2g&2-2g\\
\hline
\end{array}
\label{eq::distrEulChar}
\end{equation}
where $\Delta E=(3+g)^{-1}$ is the difference between two neighboring levels.

\subsubsection{Analogy between the ``magnetization" and the function  $f(x)=1-2h(x)$ on $\mathbb{T}_g^2$.}

It is well known that, in the context of the topological approach to PT, the energy level is a proper height function in the context of Morse theory. In the following, we will consider the Euler integral of the function, $f = 1 -2h(x)$, which we choose in analogy with the magnetization both in the $1d$- and the MF- $XY$ Models; the magnetization at a critical point $q$ is $M(q)=1-2E(q)$, where $E(q)$ is the energy at point $q$ in configuration space. 
Let us calculate $\langle 1-2h\rangle_{\chi}$, i.e.
\begin{equation}
\langle 1-2h\rangle_{\chi}(E) = \frac{\int_{M_E}(1-2h){\rm d}\chi}{\int_{M_E}{\rm d}\chi}.
\end{equation}

One has $E_0=0$, $E_k^b=k(3+g)^{-1}$, $E_k^t=(2+k)(3+g)^{-1}$, and $E_1=1$. So, for $E\in[E_i^t,E_{i+1}^t)$ and $i\in\{1,...,g-2\}$, 
the levels $E_0$, $E_1^t,\cdots,E_{i}^t$, $E_1^b,\cdots,E_{i+2}^b$ are below $E_i^t$, and it follows that
\begin{equation}
\langle 1-2h\rangle_{\chi}(E)=1+\frac{6}{(1+2i)(3+g)}-\frac{2(i+2)(i+3)}{(1+2i)(3+g).}\label{eq::meanMagaux1}
\end{equation}
The $k\mbox{-th}$ (height) level has a value of $k/(g+3)$, where $k=0,...,g+3$. The index $i$ in the expression above can be associated with the level $E_i^t$, i.e., the $(i+2)\mbox{-th}$ level. Then, we can rewrite  (\ref{eq::meanMagaux1}) in the limit $g\gg1$ as
\begin{eqnarray}
\langle 1-2h\rangle_{\chi}(E_{i+2}) 
& \stackrel{g\gg1}{=} & 1-E_{i+2}.\label{eq::Media1menosh_imenor_gm1}
\end{eqnarray}

For $E\in[E_{g-1}^t,E_{g}^t)$, the levels $E_0,E_1^t,...,E_{g-1}^t$ and $E_1^b,...,E_g^b$ are below $E_{g-1}^t$. Then,
\begin{eqnarray}
\langle 1-2h\rangle_{\chi}(E) 
& = & 1-\frac{2(2-g)}{(2-2g)(3+g)}+\frac{2g(1+g)}{(2-2g)(3+g).}\nonumber\\
\end{eqnarray}
The $(g+1)\mbox{-th}$ (height) level corresponds to the level $E_{g-1}^t=(1+g)/(3+g)$ and, therefore, we can rewrite the expression above in the limit $g\gg1$ as
\begin{eqnarray}
\langle 1-2h\rangle_{\chi}(E_{g+1}) 
& \stackrel{g\gg1}{=} & 1-E_{g+1}.\label{eq::Media1menosh_iigual_gm1}
\end{eqnarray}

On the other hand, for $E\in [E_g^t,E_1)$,  the levels $E_0,E_1^t,...,E_{g}^t$ and $E_1^b,...,E_g^b$ are below $E_{g}^t$. So,
\begin{equation}
\langle 1-2h\rangle_{\chi}(E)= 1+\frac{4g}{(1-2g)(3+g)}+\frac{2g(g+1)}{(1-2g)(3+g)}.
\end{equation}
The $(g+2)\mbox{-th}$ (height) level corresponds to the level $E_{g}^t=(2+g)/(3+g)$ and, therefore, we can rewrite the expression above in the limit $g\gg1$ as
\begin{eqnarray}
\langle1-2h\rangle_{\chi}(E_{2+g}) 
& \stackrel{g>>1}{=} & 1-E_{2+g}\label{eq::Media1menosh_iigual_g}
\end{eqnarray}

Finally, for $E\in[E_1,\infty)$, we have
\begin{equation}
\langle 1-2h\rangle_{\chi}(E)\!\!=\!\!-\frac{2g}{2-2g}+\!\frac{4g}{(2-2g)(3+g)}+\!\frac{2g(g+1)}{(2-2g)(3+g)}.
\end{equation}
The $(g+3)\mbox{-th}$ (height) level corresponds to the last level $E_1=(3+g)/(3+g)=1$ and, therefore, we can rewrite the expression above in the limit $g\gg1$ as
\begin{eqnarray}
\langle1-2h\rangle_{\chi}(E_1) 
&\stackrel{g\gg1}{=} & 1-E_1=1-1=0.\label{eq::Media1menosh_Emaximo}
\end{eqnarray}

From equations (\ref{eq::Media1menosh_imenor_gm1}), (\ref{eq::Media1menosh_iigual_gm1}), (\ref{eq::Media1menosh_iigual_g}), and (\ref{eq::Media1menosh_Emaximo}), we conclude that in the limit $g\gg1$, the mean value of the ``magnetization'' function $\langle1-2h\rangle_{\chi}(E)$ is a continuous function given by
\begin{equation}
\langle1-2h\rangle_{\chi}(E) = 1-E,\,E\in[0,1],
\end{equation}
and varies from one to zero in the energy interval.
\subsection{The constant function $f\equiv1$ on the deformed genus $g$ surface $\mathbb{T}^2_{g,\epsilon}$}
\label{sec:genusgEpsilon}

\begin{figure}[t]
\centering
\includegraphics[width=0.45\textwidth]{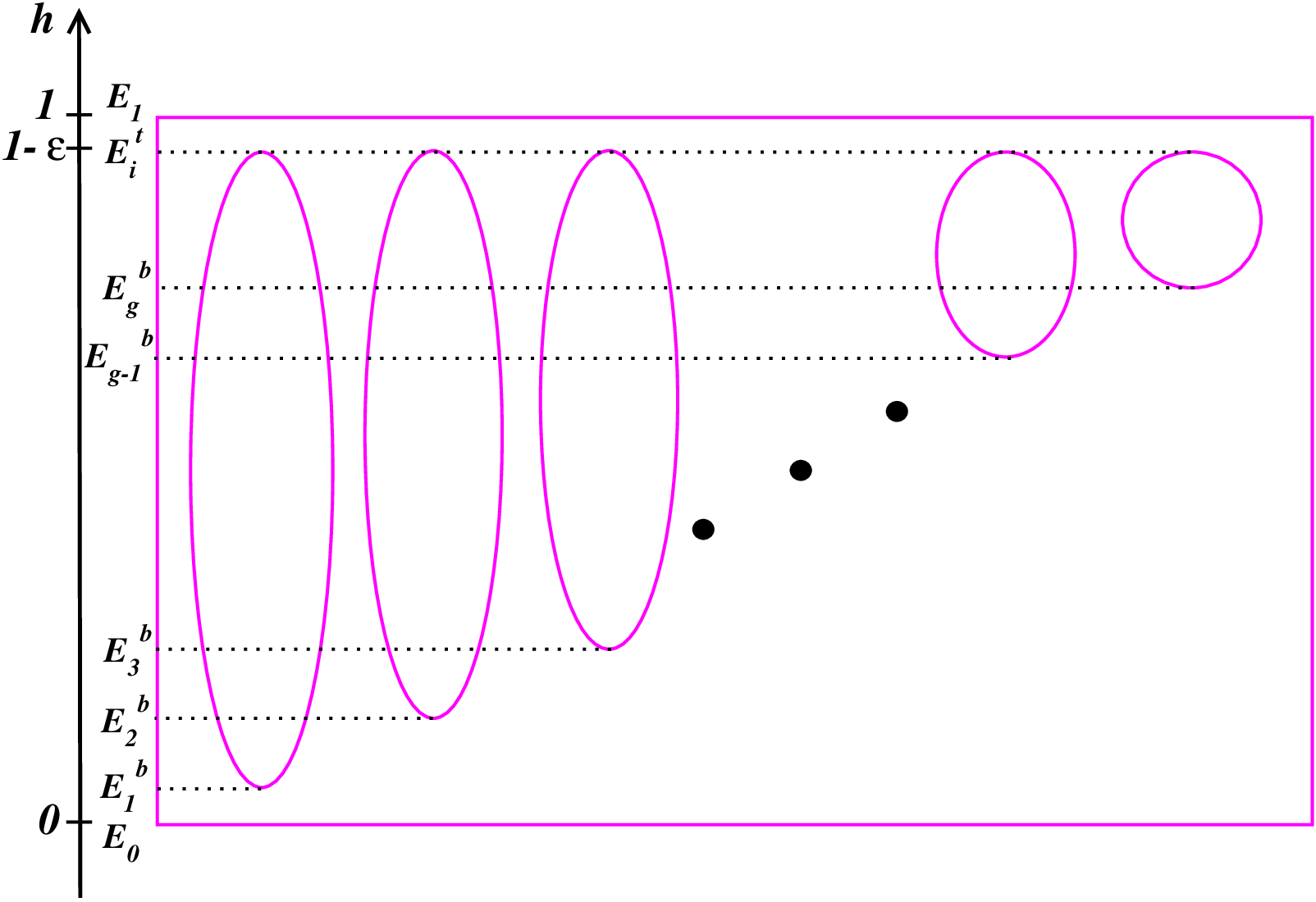}
\caption{\label{fig:T2gEpsilon}The height function $h$ on a deformed genus $g$ closed surface $\mathbb{T}_{g,\epsilon}^2$.}
\end{figure}

Now, we consider the deformed genus $g$ closed surface $\mathbb{T}_{g,\epsilon}^2$ illustrated in figure A4. Except for the last two levels, the distance between two neighboring levels (among the $g+1$ first levels) in $\mathbb{T}_{g,\epsilon}^2$ is always the same, say $\Delta E$ (figure A4). Note that each new hole adds 1 more level to the height of the surface. This implies $0=E_0<E_1^b<E_2^b<...<E_g^b<E_1^t=...=E_g^t<E_1=1$, and  a total number of $(g+3)$ levels. Again, we have $\Delta E=(3+g)^{-1}$ and, therefore,
\begin{equation}
E_k^b = \frac{k}{3+g},\,k=1,...,g;
\end{equation}
and also, due to the deformation of the top of each hole,
\begin{equation}
E_k^t = 1-\epsilon\,.
\end{equation}
Notice that those critical points corresponding to $E_i^b$ and $E_i^t$ have index 1, while those corresponding to $E_0=0$ and $E_1=1$ have index $0$ and $2$, respectively.

Let $f$ be the constant $1$ function on $\mathbb{T}^2_{g,\epsilon}.$ For $E\in[E_i^b,E_{i+1}^b)$, we have
\begin{eqnarray}
\chi(M_E)=\int_{M_E}\,{\rm d}\chi 
&=&1-i.
\end{eqnarray}
Observe that the expression above also applies to $E\in[E_g^b,E^t_i)$, i.e.
\begin{eqnarray}
\chi(M_E)=\int_{M_E}\,{\rm d}\chi & = & 1-g.
\end{eqnarray}
On the other hand, for $E\in[E_i^t,E_1),$ one has
\begin{eqnarray}
\chi(M_E)=\int_{M_E}\,{\rm d}\chi 
& = & 1-2g.
\end{eqnarray}
Finally, for $E\in[E_1,\infty)$, it follows
\begin{eqnarray}
\chi(M_E)=\int_{M_E}\,{\rm d}\chi 
& = & 2-2g,
\end{eqnarray}
which is the well known result for the Euler characteristic of an orientable closed surface of genus $g$.

In summary, the values of $\chi(M_E)$ are arranged in the table below
\begin{equation}
\begin{array}{|c|c|c|c|c|c|c|c|c|}
\hline
i\mbox{-th}\Delta E & 0 & 1 &2&3&...&g+1&g+2&g+3\\
\hline
\chi(M_E) & 1 & 0        & -1      & -2      &...& 1-g         &1-2g         &2-2g\\
\hline 
\end{array}\ ,
\label{eq::distrEulCharTepsilon}
\end{equation}
where $\Delta E=(3+g)^{-1}$ is the difference between two neighboring levels (among the first $g-1$ levels).
 
\subsection{The ``magnetization'' function $1-2h(x)$ on $\mathbb{T}^2_{g,\epsilon}$}

As already mentioned, we shall integrate the function $f(x)=1-2h(x),x\in\mathbb{T}_{g,\epsilon}^2$, in analogy with the magnetization of the 1d- and MF-XY models. For $E\in[E_i^b,E_{i+1}^b)$, we have
\begin{eqnarray}
\langle 1-2h(x) \rangle_{\chi}(E) 
& = & 1+\frac{i(i+1)}{(1-i)(3+g)}.
\end{eqnarray}
The $k\mbox{-th}$ (height) level has a value of $k/(g+3)$, where $k=0,...,g$, which corresponds to $E_k^b$ (the other levels are $E_j^t=1-\epsilon$, $\forall \,j$, and $E_1=1$). The index $i$ in the formula above can be associated with the $i\mbox{-th}$ level $E_i^b=i/(3+g)$ and, therefore, it follows in the limit $g\gg1$ that
\begin{eqnarray}
\langle1-2h(x)\rangle_{\chi}(E_i) 
& \stackrel{g\gg1}{=} & 1-E_i.\label{eq::Media1menosh_epsilon}
\end{eqnarray}
It is not difficult to see that the above expression is also valid for $E\in[E_g^b,E_i^t)$. In fact, one only needs to perform the summation from $k=1$ to $k=g$, and the  calculations that follow are similar.

On the other hand, for $E\in[E_i^t,E_1)$, we have (recall that all the $g$ points corresponding to $E_1^t,...,E_g^t$ have index $1$ and the same height $1-\epsilon$); thus
\begin{eqnarray}
\langle 1-2h(x) \rangle_{\chi}(E) 
& = & 1 +\frac{2g(1-\epsilon)}{1-2g}+\frac{g(1+g)}{(1-2g)(3+g)}\nonumber\\
& \stackrel{g\gg1}{=} & -\frac{1}{2}+\epsilon.\label{eq::Media1menosh_epsilon_AposanexarGalcas}
 \end{eqnarray}
Finally, for $E\in[E_1,\infty)$, we find
\begin{eqnarray}
\langle 1-2h(x) \rangle_{\chi}(E) 
&\stackrel{g\gg1}{=}&-\frac{1}{2}+\epsilon. \label{eq::Media1menosh_epsilon_hmax}
\end{eqnarray}

In summary, taking the limits $g\gg1$ and $\epsilon\to0$ in  (\ref{eq::Media1menosh_epsilon}), (\ref{eq::Media1menosh_epsilon_AposanexarGalcas}), and (\ref{eq::Media1menosh_epsilon_hmax}), the ``magnetization'' function $\langle1-2h(x)\rangle_{\chi}(E)$ presents a discontinuity at $E=1$:
\begin{equation}
\langle1-2h(x)\rangle_{\chi}(E) = \left\{
\begin{array}{ccc}
1-E & , & E\in[0,1)\\[7pt]
-\displaystyle\frac{1}{2} & , & E = 1\\
\end{array}
\right..
\end{equation}

\subsection{The Euler Temperature for the genus $g$ surfaces}

As seen above, the function that works as an analog for the magnetization on the genus $g$ surfaces considered as a 2-dimensional model for a ``topological transition'' behaves differently in the two cases analyzed. Indeed, for the regular surface $\mathbb{T}^2_g$ the magnetization is continuous, while for the singular one $\mathbb{T}^2_{g,\epsilon}$ the magnetization is discontinuous at the level which corresponds to the topological transition.  Now let us evaluate the Euler temperature $T_{\chi}$ for the two genus $g$ surfaces $\mathbb{T}^2_{g}$ and $\mathbb{T}^2_{g,\epsilon}$ :
\begin{equation}
\beta_{\chi} = \frac{1}{T_{\chi}} = \lim_{g\to\infty}\frac{1}{g}\frac{\ln\vert\chi(M_{E+\Delta E})\vert-\ln\vert\chi(M_{E})\vert}{\Delta E}\,.
\end{equation}
Again, the genus $g$ is seen as representing the number of degrees of freedom in the 2-dimensional model for a topological transition.

\subsubsection{$T_{\chi}$ for the regular genus $g$ surface $\mathbb{T}^2_{g}$}

Since the second value of $\chi(M_E)$ is 0, the inverse temperature $\beta_{\chi}$ shows a singularity, $\beta_{\chi}=\infty$, when we pass through this level (note that it is a trivial singularity, which can be easily removed by replacing $\vert\chi(M_{E})\vert$ by $\vert\chi(M_{E})\vert+1$ in the argument of the logarithm).

For the next $g-1$ levels, one has $\chi(M_{E_k})=-(1+2k),\,k=0,...,g-2$ and $\Delta E=(3+g)^{-1}$. Then, the inverse Euler temperature at the $(k+2)\mbox{-level}$ ($k=0,...,g-2$) is
\begin{eqnarray}
\beta_{\chi}(E_{k+2}) 
 & = & \displaystyle\frac{3+g}{g}\ln\left\vert\frac{1+2k}{1-2k}\right\vert\nonumber\,.\\
\end{eqnarray}
The height of the $(k+2)\mbox{-level}$ is $E_{k+2}=(k+2)/(3+g)$ and, therefore, in the limit $g\gg1$ one has
\begin{eqnarray}
\beta_{\chi}(E_{k+2}) 
 &\stackrel{g\gg1}{=} & 0\,.
\end{eqnarray}
For the remaining levels, the result is the same. In summary,
\begin{equation}
\beta_{\chi}(E) = \left\{
\begin{array}{ccc}
+\infty & , & E = 0\\
0 & , & E \in (0,1]\\
\end{array}
\right.\,.
\end{equation}
Note: The infinity at $E=0$ can be removed by replacing $\vert\chi(M_E)\vert$ by $\vert\chi(M_E)\vert+1$ in the argument of the logarithm.

\subsubsection{$T_{\chi}$ for the singular deformed genus $g$ surface $\mathbb{T}^2_{g,\epsilon}$}

As in the case of $\mathbb{T}^2_g$, the second value of $\chi(M_E)$ is 0 and it follows that the inverse temperature $\beta_{\chi}$ shows a singularity, $\beta_{\chi}=\infty$, when we pass through this level (similarly, it is a trivial singularity, which can be easily removed by replacing $\vert\chi(M_{E})\vert$ by $\vert\chi(M_{E})\vert+1$ in the argument of the logarithm).

For the next $g-1$ levels, one has $\chi(M_{E_k})=-k,\,k=1,...,g-1$ and $\Delta E=(3+g)^{-1}$. Then, the inverse Euler temperature at the $(k+2)\mbox{-level}$ ($k=0,...,g-2$) is
\begin{eqnarray}
\beta_{\chi}(E_{k+2}) 
 & = & \displaystyle\frac{3+g}{g}\ln\left\vert\frac{k}{1-k}\right\vert\nonumber\,.\\
\end{eqnarray}
The height of the $(k+2)\mbox{-level}$ is $E_{k+2}=(k+2)/(3+g)$ and, therefore, in the limit $g\gg1$ one has
\begin{eqnarray}
\beta_{\chi}(E_{k+2}) 
 &\stackrel{g\gg1}{=} & 0\,.
\end{eqnarray}
On the other hand, when passing through the last three levels one has $\Delta E = [2/(3+g)-\epsilon]$ and $\Delta E = \epsilon$. Therefore, in the limit $g\gg1$
\begin{eqnarray}
\beta_{\chi}(E_{g+2}) 
 & \stackrel{g\gg1}{=} & \frac{1}{2-\epsilon g}\ln2
\end{eqnarray}
and
\begin{eqnarray}
\beta_{\chi}(E_{g+3}) 
 & = & \displaystyle\frac{1}{\epsilon g}\ln\left[\frac{2g-2}{2g-1}\right]\nonumber\,.\\
\end{eqnarray}
Here, the result depends crucially on the parameter $\epsilon$. We can assume that $\epsilon=1/g^2$, by analogy with the MF-XY model, where the distance between neighboring levels changes from $1/N$ to $1/N^2$ at the critical point [See (\ref{delta})]. Under this hypothesis, one finds
\begin{equation}
\beta_{\chi}(E=1^{-}) = \ln2/2;\, \beta_{\chi}(E=1)=-\infty\,.
\end{equation}
In summary,
\begin{equation}
\beta_{\chi}(E)=\left\{
\begin{array}{ccc}
+\infty & , & E = 0\\
0 & , & E \in (0,1)\\[4pt]
\frac{1}{2}\ln2 & , & E=1^{-}\\
-\infty & , & E = 1\\
\end{array}
\right.\,.
\end{equation}
Note: The infinity at $E=0$ can be removed by replacing $\vert\chi(M_E)\vert$ by $\vert\chi(M_E)\vert+1$ in the argument of the logarithm. On the other hand, the same is not true for the singularity at $E=1$.


\section*{References}

\begin{thebibliography}{10}

\bibitem{Pettinilivro} 
Pettini M 2007 {\it Geometry and Topology in Hamiltonian Dynamics
  and Statistical Mechanics} (New York: Springer).

\bibitem{KastnerRMP2008}
Kastner M 2008 {\it Rev. Mod. Phys.} {\bf 80} 167

\bibitem{FranzosiPRL2004} 
Franzosi R and Pettini M 2004 {\it Phys. Rev. Lett.} {\bf 92} 060601\\
Franzosi R, Pettini M and Spinelli L 2007 {\it Nucl. Phys.} B {\bf 782} 189\\
Franzosi R and Pettini M 2007 {\it Nucl. Phys.} B {\bf782} 219

\bibitem{KastnerPRL2007} 
Kastner M, Schreiber S and Schnetz O 2007 {\it Phys. Rev. Lett.} {\bf
  99} 050601\\  
  Kastner M and Schnetz O 2008 {\it Phys. Rev. Lett.} {\bf 100}
  160601\\
 Kastner M, Schnetz O and Schreiber S 2008 {\it J. Stat. Mech. } P04025

\bibitem{SantosPRE2009} 
Santos F A N and Coutinho-Filho M D 2009 {\it Phys. Rev.} E {\bf 80} 031123

\bibitem{AB2chains} 
For additional information on $AB_2$ chains under frustration- or field-induced PT's, see, respectively: \\
Ten\'orio A S F,  Montenegro-Filho R R and Coutinho-Filho M D 2009 {\it Phys. Rev.} B {\bf 80} 054409\\ 
  Vitoriano C, Coutinho-Filho M D and Raposo E P 2002 {\it J. Phys.} A: {\it Math. Gen.} {\bf 35} 9049

\bibitem{NardiniPRE2009}
Nardini C and Casetti L 2009 {\it Phys. Rev. }E {\bf 80} 060103

\bibitem{KastnerPRL2011}
Kastner M and Mehta D 2011 {\it Phys. Rev. Lett.} {\bf 107} 160602\\
Mehta D, Hauenstein J D and Kastner M 2012 {\it Phys. Rev.} E {\bf 85}  061103

\bibitem{FranzosiPRL2000}
Franzosi R, Pettini M and Spinelli L 2000 {\it Phys. Rev. Lett.} {\bf  84} 2774

\bibitem{NerattiniPRE2013}
Nerattini R, Kastner M, Mehta D and Casetti L 2013 {\it Phys. Rev.} E  {\bf 87} 032140

\bibitem{theo}
Donato I, Gori M, Pettini M, Petri G, De Nigris S, Franzosi R and Vaccarino F 2016 {\it Phys. Rev.} E {\bf 93} 052138\\
Gori M, Franzosi R and Pettini M, e-print arXiv:1602.01240

\bibitem{PRL1997}
Caiani L, Casetti L, Clementi C and Pettini M 1997 {\it Phys. Rev. Lett.} {\bf 79} 4361

\bibitem{RehnBJP2012}
Rehn J A, Santos F A N and Coutinho-Filho M D 2012 {\it Braz. J. Phys.} {\bf 42} 410

\bibitem{Blanchard}
 Blanchard B, Fortunato S and Gandolfo D 2002 {\it Nucl. Phys.} B {\bf  644} 495\\ 
  Blanchard P, Dobrovolny C, Gandolfo D and Ruiz J 2006 {\it J. Stat. Mech.} {\bf 03} PO3011
  
\bibitem{congressoStanley70}
Santos F A N, Rehn J A and Coutinho-Filho M D 2014 {\it Topological and Geometric Aspects of Phase Transitions}, in {\it Perspectives and Challenges in Statistical Physics and Complex Systems for the Next Decade}, ed. Viswanathan G M, Raposo E P and da Luz M G E (Singapore: World Scientific Publishing)

\bibitem{Milnor}
Milnor J 1963 {\it Morse Theory, Annals of Mathematical Studies},
  Vol. 51 (Princeton: Princeton University Press)

\bibitem{Matsumoto}
Matsumoto Y 2002 {\it An Introduction to Morse Theory}, Volume 208
  of {\it Translations of Mathematical Monographs} (Providence: American Mathematical
  Society)

  
  
\bibitem{CasettiJSP2003}
 Casetti L, Pettini M and Cohen E G D 2003 {\it J. Stat. Phys.} {\bf111} 1091, and references therein

\bibitem{AngelaniPRE2005}
Angelani L, Casetti L, Pettini M, Ruocco G and Zamponi F 2003 {\it Europhys. Lett.} {\bf 62} 775\\ 
Angelani L, Casetti L, Pettini M, Ruocco G, and Zamponi F 2005 {\it Phys. Rev.} E {\bf71} 036152

\bibitem{Kubo}
Kubo R 1965 {\it Statistical Mechanics - An Advanced Course With
  Problems and Solutions} (Amsterdam: North-Holland Publishing Company)

\bibitem{Viro}
Viro O 1988 {\it Some integral calculus based on the Euler
  characteristic}, Lecture Notes in Math. {\bf 1346}, 127-138 (New York: Springer)

\bibitem{cha}
Schapira P 1991 {\it J. Pure Appl. Algebra} {\bf 72} 83\\ 
Kashiwara M and Schapira P 1994 {\it Sheaves on Manifolds} (New York: Springer)

\bibitem{Baryshnikov}
Baryshnikov Y and Ghrist R 2010 {\it Proc. Natl. Acad. Sci.} {\bf 107} 9525
  
\bibitem{bell}
Baryshnikov Y and Ghrist R 2009 {\it SIAM J. Appl. Math.} {\bf 70} 825\\ 
for applications of the Euler characteristic to the study of robot arm, see: Farber M and Fromm V 2011 {\it J. Aust. Math. Soc.} {\bf 90} 183
  
\bibitem{Gallavotti}
Gallavotti G 1999 {\it Statistical Mechanics: A Short Treatise} (New York: Springer)
  
\bibitem{Palais}
Palais R S and Terng C 1988 {\it Critical Point Theory and
  submanifold Geometry, Lecture Notes in Mathematics}, {\bf 1353} (Berlin: Springer)

\bibitem{AndronicoPRE2004}
Andronico A, Angelani L, Ruocco G and Zamponi F 2004 {\it Phys. Rev.} E {\bf70} 041101

\bibitem{GrinzaPRL2004}
Grinza P and Mossa A 2004 {\it Phys. Rev. Lett.} {\bf92} 158102


\bibitem{AngelaniPRE20051dModels}
Kastner M 2004 {\it Phys. Rev. Lett.} {\bf93} 150601\\
Angelani L, Ruocco G and Zamponi F 2005 {\it Phys. Rev.} E {\bf72} 016122

\bibitem{BraunScience2013}
Braun S, Ronzheimer J P, Schreiber M, Hodgman S S, Rom T,  Bloch I and Schneider U 2013 {\it Science} {\bf 339} 52 
  
\bibitem{DunkelNaturePhys2014}
Dunkel J and Hilbert S 2014 {\it Nature Phys.} {\bf 10} 67\\
Dunkel J and Hilbert S 2006 {\it Physica} A {\bf 370} 390
  
\bibitem{RuffoPRE1995}
Antoni M and Ruffo S 1995 {\it Phys. Rev.} E {\bf 52} 2361\\
Campa A, Dauxois T and Ruffo S 2009 {\it Phys. Rep.} {\bf 480} 57
  
\bibitem{Mehta}
Mehta D and Kastner M 2011 {\it Ann. Phys.} {\bf 326} 1425
  
\bibitem{khinchinANDkubo}
Khinchin A I 1960 {\it Mathematical Foundations of Statistical
  Mechanics} (New York: Dover Publications)\\
  Khinchin A Y 1998 {\it
  Mathematical Foundations of Quantum Statistics} (New York: Dover Publications)

\bibitem{purcel}
Purcell E M and Pound R V 1951 {\it Phys. Rev.} {\bf 81} 279

\bibitem{ramsey}
Ramsey N F 1956 {\it Phys. Rev.} {\bf 103} 20

\bibitem{klein}
Klein M J 1956 {\it Phys. Rev.} {\bf 104} 589

\bibitem{ShenScripta2003}
Shen J-Q 2003 {\it Phys. Scr.} {\bf 68} 87

\bibitem{G-DPRD2004}
Gonzalez-Diaz P F 2004 {\it Phys. Rev.} D {\bf 70} 063530

\bibitem{PariharPRC2006}
Parihar V, Widom A and Srivastava Y N 2006 {\it Phys. Rev.} C {\bf  73} 017901

\bibitem{SchmidtPRE2005}
Schmidt H and Mahler G 2005 {\it Phys. Rev.} E {\bf 72} 016117

\bibitem{ParisiPRA2010}
De Pasquale A, Facchi P, Parisi G, Pascazio S and
  Scardicchio A S 2010 {\it Phys. Rev.} A {\bf 81} 052324
  
\bibitem{MoskPRL2005}
Mosk A P 2005 {\it Phys. Rev. Lett.} {\bf 95} 040403\\
Rapp A,
  Mandt S and Rosch A 2010 {\it Phys. Rev. Lett.} {\bf 105} 220405\\
  Tsuji N, Oka T, Werner P and Aoki H 2011 {\it Phys. Rev. Lett.} {\bf 106} 236401 

\bibitem{RomeroRochinPRE2013}
Romero-Roch\'in V 2013 {\it Phys. Rev.} E {\bf 88} 022144

\bibitem{LebowitzPhysA}
Lebowitz J L 1993 {\it Physica} A {\bf 194} 1\\
Lebowitz J L 1999 {\it Rev. Mod. Phys.} {\bf 71} S346

\bibitem{gross}
Gross D H E 2001 {\it Microcanonical Thermodynamics - Phase
  Transitions in ``Small'' Systems} (Singapore: World Scientific)\\
Lebowitz J L and Percus J K 1961 {\it Phys. Rev.} {\bf 124} 1673

\bibitem{SantosFigE}
See figure 7(c) of \cite{SantosPRE2009}

\bibitem{footnoteSantos}
These three distinct solutions were found for $T \rightarrow 0$
  in  \cite{SantosPRE2009}.

\bibitem{griff}
Griffthis R B, Weng C-Y and Langer J S 1966 {\it Phys. Rev.} {\bf
  149} 301\\
O. Penrose O and Lebowitz J L 1971 {\it J. Stat. Phys.} {\bf
  3} 211

\bibitem{Chaikin} 
A general analysis of the dynamics of a variety of systems, including nucleation and spinodal decomposition, can be found in: Chaikin P M and Lubensky T C 1995 {\it Principles of Condensed Matter Physics} (Cambridge: Cambridge University Press), Chapter 8

\bibitem{ParisiMeta}
Parisi G 1988 {\it Statistical Field Theory} (Redwood City, California: Addison-Wesley
  Publishing Company Inc.), p. 201
  
\bibitem{Fisher}
Fisher M E and Zinn S-Y 1998 {\it J. Phys.} A:{\it Math. Gen.} {\bf 31}
  L629\\
  Zinn S-Y and Fisher M E 2005 {\it Phys. Rev.} E {\bf 71} 011601, and references therein

\bibitem{LimitBinomial}
Using the relation
  between $n_{\pi}^{(-)}$ and $n_{\pi}^{(+)},$ we can write $X=\frac{{N\choose
  N-n_{\pi}^{(+)}}}{{N\choose
  n_{\pi}^{(-)}}}=\frac{(n_{\pi}^{(-)})(n_{\pi}^{(-)}-1)\cdots(n_{\pi}^{(-)}-Nh+1)}{(n_{\pi}^{(+)}-1)(n_{\pi}^{(+)}-2)\cdots(n_{\pi}^{(+)}-Nh+1)}.$
  Noting that there are the same quantity of factors at the numerator and
  denominator, we have $\ln X =-Nh(2N\sqrt{h^2-2(E-\frac{1}{2})}).$ Finally, taking the limit
  $N\to\infty,$ it results that $\ln X\to-\mbox{sign}(h)\times\infty$ and the desired result
  follows

\bibitem{LuizkTM}
Da Silva L C B, Santos F A N and Coutinho-Filho M D 2013 unpublished  
  
\bibitem{StanleyXY1d}
Stanley H E 1969 {\it Phys. Rev.} {\bf 179} 570\\
Thompson C J 1988 {\it Classical Equilibrium Statistical Mechanics} (New York: Oxford University Press)

\bibitem{DenisovPRE1993}
Denisov S I and Hanggi P 2005 {\it Phys. Rev.} E {\bf 71} 046137

\bibitem{FisherAJP1964}
Fisher M E 1964 {\it Am. J. Phys.} {\bf 32} 343\\
  Blume M and Heller P 1975 {\it Phys. Rev.} B {\bf 11} 4483
  
 \bibitem{Livro} 
 Ghrist R 2014 {\it Elementary Applied Topology} (CreateSpace Independent Publishing Platform)
 
\bibitem{LivroShape}  
Patrikalakis N M and Maekawa T 2010 {\it  Shape Interrogation for Computer Aided Design and Manufacturing} (Berlin: Springer)

\end{thebibliography}
\end{document}